\documentclass[a4paper,11pt]{article}

\pdfoutput=1

\usepackage{jheppub}
\usepackage[T1]{fontenc} 
\usepackage{graphicx}
\usepackage{epsfig}
\usepackage{rotating}
\usepackage{amssymb}
\usepackage{subfigure}
\usepackage{dsfont}
\usepackage{psfrag}
\usepackage{amsmath,euscript,array,mathrsfs}
\usepackage{bbold}
\usepackage{epsf}
\usepackage{slashed}

\usepackage{tikz}
\usetikzlibrary{decorations.markings,decorations.pathmorphing}

\newcommand{\beq}{\begin{equation}}
\newcommand{\eeq}{\end{equation}}
\newcommand{\beqs}{\begin{eqnarray}}
\newcommand{\eeqs}{\end{eqnarray}}

\newcommand{\dd}{\mathrm{d}}

\title{Mass spectrum of gapped, non-confining theories with multi-scale dynamics}

\author[1,2]{Daniel Elander,}
\author[2]{Ant\'on F. Faedo,}
\author[2,3]{David Mateos,}
\author[2]{David Pravos,}
\author[2]{Javier G. Subils}

\affiliation[1]{Laboratoire Charles Coulomb (L2C), University of Montpellier, CNRS, Montpellier, France}
\affiliation[2]{Departament de F\'isica Qu\`antica i Astrof\'isica \& Institut de Ci\`encies del Cosmos (ICC), \\ Universitat de Barcelona, Mart\'i Franqu\`es 1, ES-08028, Barcelona, Spain}
\affiliation[3]{Instituci\'o Catalana de Recerca i Estudis Avan\c cats (ICREA), Passeig Llu\'\i s Companys 23, \\ ES-08010, Barcelona, Spain}

\date{\today}

\abstract{
We study the mass spectrum of spin-0 and spin-2 composite states in a one-parameter family of three-dimensional field theories by making use of their dual descriptions in terms of supergravity. These theories exhibit a mass gap despite being non-confining, and by varying a parameter can be made to flow arbitrarily close to an IR fixed point corresponding to the Ooguri--Park  conformal field theory. At the opposite end of parameter space, the dynamics becomes quasi-confining. The glueball spectrum interpolates between these two limiting cases, and for nearly conformal dynamics approaches the result of the Ooguri--Park theory deformed by a relevant operator. In order to elucidate under which circumstances quasi-conformal dynamics leads to the presence of a light pseudo-dilaton, we perform a study of the dependence of the spectrum on the position of a hard-wall IR cutoff and find that, in the present case, the mass of such state is lifted by deep-IR effects.
}

\begin{document}

\begin{flushright}
\hfill{ICCUB-18-019}
\end{flushright}

\maketitle
\flushbottom

\section{Introduction}

Strongly coupled field theories in which more than one characteristic energy scale is dynamically generated allow for a rich set of phenomena due to observables developing dependences on ratios of such scales. Within the framework of gauge-gravity duality \cite{AdSCFT}, there exists a number of examples constructed in string theory that exhibit non-trivial renormalization group (RG) flows leading to multi-scale dynamics \cite{MultiScale}. The reformulation of strongly coupled gauge theories in terms of classical higher-dimensional gravity provides a powerful tool with which analytical progress can be made.

One way to obtain strongly coupled field theories with multi-scale dynamics is to consider cases where the RG flow comes close to an IR fixed point, leading to quasi-conformal (walking) behaviour over a range of energies. Such walking theories have been proposed as phenomenological models in which the electro-weak symmetry is broken by strong coupling effects and the big and little hierarchy problems are solved dynamically without the need for fine-tuning \cite{WTC}. A crucial question is under which circumstances the existence of a walking region leads to a parametrically light pseudo-dilaton in the spectrum, due to the spontaneous breaking of approximate scale invariance. In the context of phenomenology such composite state may be identified with the 125 GeV Higgs boson discovered at the Large Hadron Collider \cite{Higgs}.

Using gauge-gravity duality, spectra of composite states in the field theory can be found by studying fluctuations around backgrounds in the dual supergravity description. When the supergravity admits a consistent truncation to a sigma model composed of a number of scalar fields coupled to gravity in $d+1$ dimensions (where $d$ is the number of field theory dimensions), there exists a powerful gauge-invariant formalism to treat the fluctuations in the bulk developed in \cite{BPM,BHM1,BHM2,ESQCD,LightScalars}, allowing for the calculation of spin-0 and spin-2 glueball spectra. This has been used to compute spectra for a number of backgrounds with walking dynamics obtained as deformations of the Maldacena--Nunez \cite{CVMN} and Klebanov--Strassler \cite{KS} backgrounds, leading to the identification of a light state when there is an operator that acquires a VEV that is parametrically larger than the scale of confinement \cite{GlueballSpectra}. Walking dynamics has also been explored holographically in a top-down context in  \cite{PremKumar:2010as,Anguelova}.

In this paper, we compute the spectrum of spin-0 and spin-2 glueballs in a class of three-dimensional gauge theories that are dual to a one-parameter family of solutions in M-theory \cite{Cvetic:2001ye,Cvetic:2001pga}. In terms of ten-dimensional type-IIA supergravity, the asymptotic UV behaviour of the solutions corresponds to placing a stack of D2-branes at a cone over $\mathbb{CP}^3$ and turning on additional fluxes. The dual is expected to be a quiver-type Chern--Simons Matter (CSM) gauge theory with gauge group $U(N)_k \times U(N+M)_{-k}$, with $k$ the Chern--Simons level. From the eleven-dimensional point of view, the solutions are all regular in the IR with an end-of-space leading to a mass gap. Despite this, the geometries describe non-confining field theories \cite{Faedo:2017fbv}, as we will review in Section~\ref{sec:summary}. 

The parameter $\tau_*$ labelling the backgrounds controls, in the gauge dual, the difference between the microscopic couplings of each of the two factors in the gauge group \cite{Hashimoto:2010bq}. By varying its value, it is possible to construct RG flows that come arbitrarily close to the IR fixed point given by the Ooguri--Park (OP) conformal field theory (CFT) \cite{Ooguri}, hence exhibiting quasi-conformal dynamics over a range of energies. In the opposite limit, the solutions are quasi-confining, coming close to a confining solution \cite{Cvetic:2001ma} denoted as $\mathbb B_8^{\rm conf}$. Moreover, the system admits a consistent truncation to a sigma model composed of six scalar fields coupled to gravity in four dimensions, allowing us to make use of the aforementioned gauge-invariant formalism to compute the spectrum as a function of $\tau_*$.

In order to gain insight into the physics underlying the various features of the spectrum, we also perform a study in which we introduce a hard-wall cutoff in the IR. Varying the energy scale at which the geometry is thus cut off, we are able to interpolate between the results obtained from a hard-wall IR cutoff and a smooth end of the geometry. We find that there are different regions of parameter space for which there is a light dilaton in the spectrum, and that, in the case of the geometry ending smoothly, the mass of such state is lifted by deep-IR effects.

The paper is organized as follows. In Section~\ref{sec:summary}, we summarize the type-IIA supergravity and M-theory solutions and their description in terms of a four-dimensional sigma model, as well as review the gauge-invariant formalism for the computation of spectra. Section~\ref{sec:spectrum} contains the numerical computation of the spin-0 and spin-2 glueball spectra. Finally, we conclude with a discussion of our results in Section~\ref{sec:conclusions}.

\section{Summary of known results}
\label{sec:summary}

\subsection{Type IIA/M-theory solutions and their field theory duals}
\label{sec:lift}

In this section we present some basic features of the ten and eleven-dimensional supergravity solutions we will consider together with their putative gauge-theory dual interpretation. For the technical details we refer to \cite{Faedo:2017fbv}. The starting point is a stack of $N$ D2-branes at the tip of a cone over $\mathbb{CP}^3$, supported by a Ramond--Ramond (RR) four-form flux proportional to $N$.\footnote{This $\mathbb{CP}^3$ is not endowed with the usual Fubini--Study metric, but with a different Einstein metric admitting a Nearly K\"ahler structure.} In the decoupling limit, this solution is expected to be dual to a three-dimensional quiver-type Yang--Mills theory with gauge group $U(N) \times U(N)$ preserving minimal supersymmetry \cite{Loewy:2002hu}. 

The backgrounds considered in this paper asymptote to this metric in the UV.\footnote{With the exception of what we call $\mathbb{B}_8^\mathrm{OP}$, which is AdS in the UV.} The internal manifold, $\mathbb{CP}^3$, can be seen as an $S^2$ fibration over $S^4$, so the D2-brane solution can be generalized by allowing the relative size of the fiber and the base to change with the radial coordinate and consequently along the RG flow of the dual theory. In particular, solutions exist in which the size of $S^4$ inside $\mathbb{CP}^3$ remains finite at the end of the geometry, that is, the deep IR, providing an additional scale in the gauge theory and pointing towards a mass gap. However, the ten-dimensional metric is singular in the IR. To be able to interpret these geometries as duals for gauge theories we need to regularize them. 

The first step is to include additional two- and four-form RR fluxes as well as turning on the NS three-form. The two-form is proportional to the K\"ahler form of $\mathbb{CP}^3$, as in ABJM \cite{Aharony}, so it will induce Chern--Simons (CS) interactions in the gauge theory dual. The new three- and four-form components can be seen to be sourced by fractional D2-branes \cite{Hashimoto:2010bq}, so similarly to \cite{KS} we expect a shift in the rank of one of the gauge groups proportional to the number $M$ of such fractional branes. Thus, the conjectured dual would be an $\mathcal{N}=1$ quiver-type Yang--Mills theory with gauge group $U(N)_k\times U(N+M)_{-k}$, with $k$ the level of the supplementary CS interactions.   

The solutions with all these features are still singular in ten dimensions, so a further uplift to eleven-dimensional supergravity is needed. The non-vanishing type-IIA two-form induces a non-trivial fibering of the M-theory circle that combines with the $S^2$ to produce a (squashed) three-sphere. At the same time, this $S^3$ is fibered over the remaining internal four-sphere to give a (squashed) seven-sphere. The eleven-dimensional geometry obtained in this way corresponds to M2-branes in the Spin(7) holonomy manifolds $\mathbb{B}_8^{\pm}$, first found in \cite{Cvetic:2001ye,Cvetic:2001pga}, and is perfectly regular. These special holonomy manifolds appear in a one-parameter family. We denote this parameter by $y_0$. As argued in \cite{Hashimoto:2010bq,Faedo:2017fbv}, we expect it to control the difference between the microscopic couplings of each of the two factors in the gauge group.

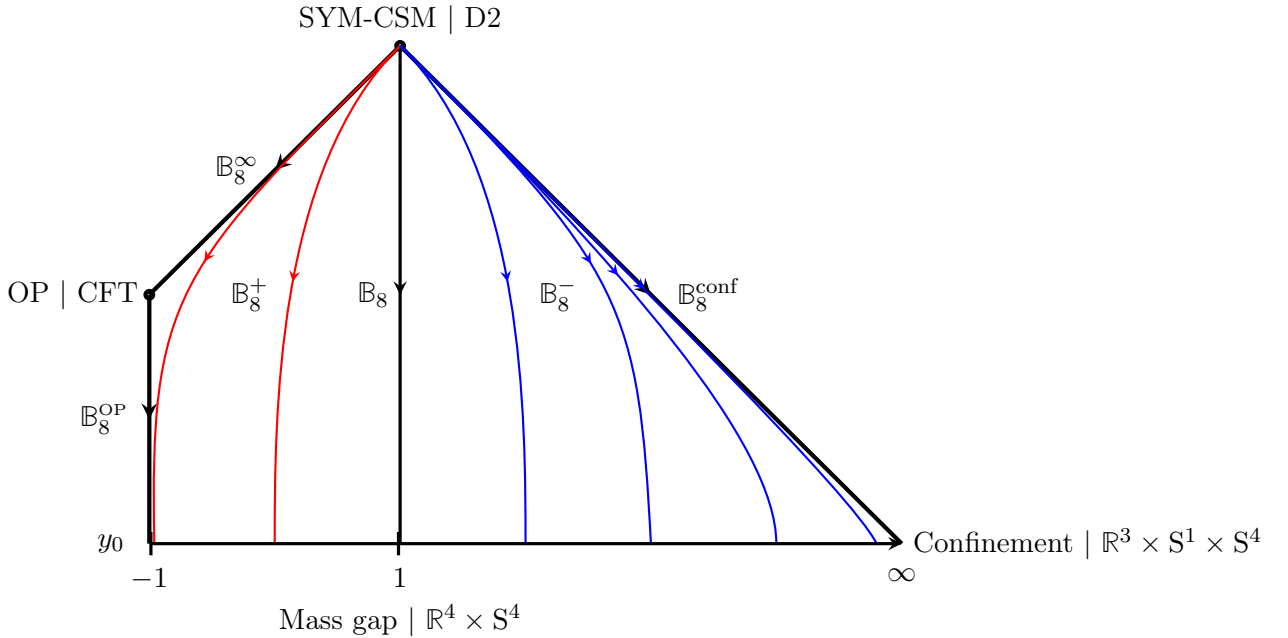
\begin{figure}[t]
\centering
\begin{tikzpicture}[scale=3.3,very thick,decoration={markings,mark=at position .5 with {\arrow{stealth}}}]
	\node[above] at (0,0) {SYM-CSM $|$ D2};
	\node[below] at (0,-2.2) {Mass gap $|$ $\mathbb{R}^4\times {\rm S}^4$};
	\node[left] at (-1,-1) {OP $|$ CFT};
	\node[right] at (2,-2) {Confinement $|$ $\mathbb{R}^3\times{\rm S}^1\times {\rm S}^4$};
	\draw [black, ultra thick] (0,0) circle [radius=0.015];
	\draw [black, ultra thick] (-1,-1) circle [radius=0.015]; 
	\node at (-.6,-1) {$\mathbb{B}_8^+$};
	\node at (.63,-1) {$\mathbb{B}_8^-$};
	\draw[postaction={decorate},very thick] (0,0) --  (0,-2) node[left,midway]{$\mathbb{B}_8$};
	\draw[postaction={decorate},ultra thick] (0,0) -- (-1,-1) node[left,midway]{$\mathbb{B}_8^\infty$\,};
	\draw[postaction={decorate},ultra thick] (0,0) -- (2,-2) node[right,midway]{\, $\mathbb{B}_8^{\rm{conf}}$};
	\draw[postaction={decorate},ultra thick] (-1,-1) -- (-1,-2) node[left,midway]{$\mathbb{B}_8^{\textrm{\tiny OP}}$\,\,};
        \draw[postaction={decorate},thick, red] (0,0) .. controls (-.9,-.9) and (-1,-1) .. (-0.98,-2);
	\draw[postaction={decorate},thick, red] (0,0) .. controls (-.5,-.5) and (-.5,-1.5) .. (-.5,-2);
	\draw[postaction={decorate},thick, blue] (0,0) .. controls (.5,-.5) and (.5,-1.5) .. (.5,-2);
	\draw[postaction={decorate},thick, blue] (0,0) .. controls (.9,-.9) and (.95,-1.05) .. (1,-2);
	\draw[postaction={decorate},thick, blue] (0,0) .. controls (.9,-.9) and (1.5,-1.6) .. (1.5,-2);
	\draw[postaction={decorate},thick, blue] (0,0) .. controls (.95,-.95) and (1.8,-1.8) .. (1.9,-2);
	\draw[|-|] (-1,-2) -- (0,-2);
	\draw[-stealth] (0,-2) -- (2,-2);
	\node[left=5] at (-1,-2) {$y_0$};
	\node[below=5] at (-1,-2) {$-1$};
	\node[below=5] at (0,-2) {$1$};
	\node[below=5] at (2,-2) {$\infty$};
	\end{tikzpicture}
\caption{\small Pictorial representation of the different solutions described in the main text. We indicate the geometry, together with the dual interpretation, as a function of the parameter $y_0$. The arrows indicate the direction of the RG flow from UV to IR. Figure taken from \cite{Faedo:2017fbv}.
}\label{fig:triangle}
\end{figure}

\begin{figure}[t]
\begin{center}
\includegraphics[width=.95\textwidth]{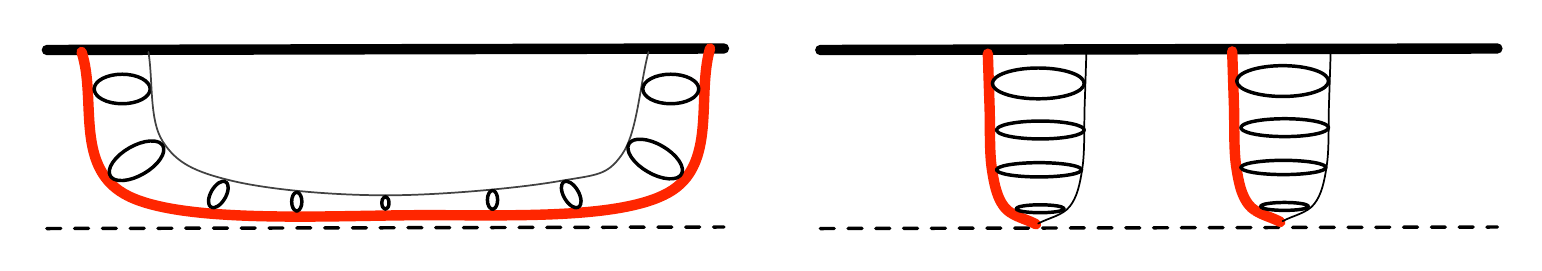} 
\caption{\small \small (Left) Connected membrane configuration in the M-theory computation of the quark-antiquark potential. The thick, red curve represents the projection onto the non-compact directions. The black circles at each point illustrate the M-theory circle. The top, horizontal line is the boundary where the gauge theory is defined. The bottom, dashed line is where the metric ends smoothly. (Right) Charge conservation allows the disconnected configuration, since the membrane closes off smoothly and therefore it has a boundaryless, cigar-like topology. Figure taken from \cite{Faedo:2017fbv}.}
\label{membrane}
\end{center}
\end{figure}

In Figure \ref{fig:triangle} we represent pictorially the set of solutions as a function of such parameter, which ranges from $-1$ to $+\infty$, together with the dual interpretation. The gauge theories exhibit different physics varying continuously with $y_0$. This can be understood in terms of the geometry of the supergravity solutions. For the values $y_0\in(-1,+\infty)$, describing the solutions $\mathbb{B}_8^{\pm}$ (and the special case $\mathbb{B}_8$ for which $y_0 = 1$), the $S^3$ shrinks to zero size smoothly in the IR, whereas the size of the $S^4$ remains finite. The IR transverse geometry is thus $\mathbb{R}^4\times S^4$. In the dual theories, this implies the existence of a mass gap without confinement. The argument for this is that a quark-antiquark pair is represented by a couple of membranes wrapped around the M-theory circle. This $S^1\subset S^3$ shrinks smoothly in the IR, so that each independent membrane can extend from the boundary to the bottom of the geometry ending in a cigar-like shape, as depicted to the right in Figure~\ref{membrane}. As the separation between the quark-antiquark pair is made large, this configuration is preferred over the connected one to the left in Figure~\ref{membrane}.

When the parameter takes the value $y_0 = -1$, the physics is completely different. The whole $S^7$ shrinks to zero size in the IR (solution $\mathbb{B}^\infty_8$), but when taking the warp factor into account, the IR geometry becomes $\mathrm{AdS}_4$ times a squashed seven-sphere of finite size. This fixed point is dual to the OP CFT \cite{Ooguri}. The OP fixed point admits a relevant deformation ($\mathbb{B}_8^\mathrm{OP}$) that drives it to an IR with the transverse geometry $\mathbb{R}^4\times S^4$ as in the previous case, leading again to a mass gap but no confinement. Solutions with $y_0$ close to $-1$ describe RG flows that approach the concatenation of the $\mathbb{B}_8^\infty$ and the $\mathbb{B}_8^\mathrm{OP}$ flows. These solutions exhibit quasi-conformal (walking) dynamics, remaining close to the OP fixed point over a tuneable range of energies.

In the case of vanishing RR two-form, one obtains a solution based on an internal geometry found in  \cite{Bryant:1989mv,Gibbons:1989er} that we call $\mathbb{B}_8^\mathrm{conf}$. The eleven-dimensional solution flows to an IR theory that exhibits both a mass gap and confinement. The geometric reason is that, in this case, the M-theory circle is trivially fibered over the rest of the geometry and it remains non-contractible along the entire flow; in particular, the IR transverse geometry is $\mathbb{R}^3\times S^1\times S^4$. This implies that a membrane wrapped on this $S^1$ cannot end anywhere in the bulk since it would have a cylinder-like geometry and hence a boundary, which is not allowed by charge conservation. The disconnected, deconfined configuration is thus not admissible. On the gauge theory side, the existence of confinement appears to be a consequence of the absence of CSM interactions \cite{Faedo:2017fbv}. This solution can also be recovered as a limit of the $\mathbb{B}_8^-$ solutions when $y_0\to+\infty$, implying that for large values of $y_0$ the solutions exhibit quasi-confining behavior.

Finally, it will be convenient to work with the redefined parameter 
\beq
\tau_*=\frac12\log\left(1+y_0\right),
\eeq
whose range is $\tau_*\in\left(-\infty,\infty\right)$. The physical results in the next sections are presented in terms of it.

\subsection{Four-dimensional description}
\label{sec:backgrounds}

Holographic spectra are conveniently computed using the gauge-invariant formalism discussed in \cite{BHM1,LightScalars}. This requires the existence of a lower-dimensional sigma model containing at least all the modes that are excited in the background solutions. Moreover, it must be a consistent truncation of ten or eleven-dimensional supergravity in order to ensure that the set of fields that one retains is closed, that is, they do not source additional modes outside the truncation.

Fortunately, all the backgrounds of interest can be obtained as a solution to a four-dimensional supergravity. The details of the full reduction from ten dimensions can be found in \cite{Cassani:2009ck}, from which we will keep only the scalars and follow the notation of \cite{Faedo:2017fbv}. The resulting sigma model contains gravity plus six scalars, three of which,  $\{ \Phi, U, V\}$, come from the metric together with the dilaton, while the rest, $\{a_J, b_J, b_X \}$, descend from the forms. The four-dimensional action is given by
\beqs
\label{eq:4daction}
	\mathcal S_4 &=& \int \dd\rho \, \dd^3 x \, \sqrt{-g}\left(\frac{R}{4}-\frac{1}{2}G_{ab} (\Phi^a) g^{MN} \partial_M\Phi^a\partial_N\Phi^b- \mathcal V(\Phi^a) \right) ,
\eeqs
where $g_{MN}$ is the four-dimensional metric ($M,N = 0,1,2,3$), $G_{ab}$ is the sigma-model metric ($a,b = 1, \cdots, 6$), whose explicit form is given by
\beqs
\label{eq:sigmametric}
	G_{ab}\partial_M\Phi^a\partial_N\Phi^b &=& \frac{1}{4} \partial_M \Phi \partial_N \Phi +2 \partial_M U \partial_N U + 6 \partial_M V \partial_N V + 4  \partial_M U \partial_N V  \\[2mm] &+& 16e^{-2U-4V+\frac{\Phi}{2}} \partial_M a_J \partial_N a_J + 2e^{-4V-\Phi} \left( \partial_M b_J \partial_N b_J  + 2 \partial_M b_X \partial_N b_X \right) ,\nonumber
\eeqs
and the potential $\mathcal V$ can be written in terms of a superpotential $\mathcal W$ as follows ($\mathcal W^a = G^{ab} \mathcal W_b = G^{ab} \frac{\partial \mathcal W}{\partial \Phi^b}$)
\beqs
 \label{eq:superpotential}
	\mathcal V &=& \frac{1}{2} \mathcal W_a \mathcal W^a - \frac{3}{2} \mathcal W^2 , \nonumber\\[2mm]
	\mathcal W &=& Q_k e^{-U-4 V+\frac{3 \Phi }{4}} - Q_k e^{-3 U-2 V+\frac{3 \Phi }{4}} - 2 e^{-2 U-2 V}-e^{-4 V} \\[2mm] &+& 4 e^{-3 U-6 V-\frac{\Phi }{4}} \Big(4 a_J \left(b_J+b_X\right)+2 q_c \left(b_X-b_J\right)+b_J Q_k \left(b_J-2 b_X\right)+Q_c \Big) .\nonumber
\eeqs
The constant parameters $Q_k$, $Q_c$ and $q_c$ appearing in the potential are two- and four-form charges. They are proportional respectively to the CS level $k$, the number of colours $N$ and the shift in the rank of one of the gauge groups $M$ due to the fractional branes. 

All the solutions preserve Poincar\'e invariance in three dimensions, so we restrict ourselves to backgrounds that only depend on a radial coordinate $\rho$, for which the metric has the form of a domain wall
\beq
	\dd s_4^2 = \dd\rho^2 + e^{2A(\rho)} \dd x_{1,2}^2 .
\eeq
Moreover, they are $\mathcal{N}=1$ supersymmetric, so as usual they can be obtained from a set of BPS equations that read
\beq\label{eq:firstordereqs}
	 \Phi'^a =G^{ab} \frac{\partial \mathcal W}{\partial \Phi^b}, \hspace{1.5cm} A' = -\mathcal W ,
\eeq
where prime denotes derivatives with respect to $\rho$. It can be seen that the equations for $U$, $V$, $\Phi$ and $A$ found in this way decouple from the rest. It is convenient to write the scalars in terms of new functions $H(y)$, $P(y)$ and $v(y)$ as\footnote{Note that for all backgrounds in this section $Q_k < 0$.}
\beqs
\begin{array}{rclcrcl}
	\Phi &=& \frac{3}{4} \log \left(\frac{4 {H^{1/3}} P (v-2)}{v^3 (y+1) Q_k^2}\right) ,&\qquad\qquad&U &=& \frac{9}{16} \log \left(\frac{4 {H^{1/3}} P (v-2)}{v^{11/9} (y+1) |Q_k|^{2/9}}\right) ,\\[3mm]
	V &=& \frac{1}{16} \log \left(\frac{1024 H^3 P^9 (v-2)}{v^3 (y+1) Q_k^2}\right) , &\qquad\qquad&A& =& \frac{3}{4} \log \left(\frac{2^{1/3} 8 {H^{1/3}} P^{7/3} (v-2)}{v^{5/3} (y+1) |Q_k|^{2/3}}\right) ,
\end{array}	
\eeqs
and change the radial coordinate from $\rho$ to $y$ defined as
\beq
	\dd\rho = \dd y \, \frac{2 \sqrt{2} H^{3/4} P^{9/4} (v-2)^{1/4}}{|Q_k|^{1/2} v^{3/4} (1-y) (y+1)^{5/4}} .
\eeq
As a consequence, the BPS equations for $\Phi$, $U$, and $V$ following from Eq.~\eqref{eq:firstordereqs} are solved provided
\beq
	\frac{\partial_y P}{P} = \frac{v+1}{v \left(1- y ^2\right)} , \qquad\qquad
	\partial_y v = \frac{v y+2}{2 \left(1-y^2\right)} .
\eeq
After the shifts and rescalings of the fluxes 
\beqs
\label{eq:rescalings}
\begin{array}{rclcrcl}
	a_J &=& -\frac{q_c}{6} + \sqrt{4q_c^2 - 3 Q_c Q_k} \, \mathcal A_J , &\qquad\qquad&	b_J& =& \frac{2q_c}{3 Q_k} + \frac{\sqrt{4q_c^2 - 3 Q_c Q_k}}{3Q_k} \mathcal B_J ,\\[3mm]
	b_X &=& -\frac{2q_c}{3 Q_k} - \frac{\sqrt{4q_c^2 - 3 Q_c Q_k}}{3Q_k} \mathcal B_X , &\qquad\qquad&H& =& \frac{4q_c^2 - 3 Q_c Q_k}{P_0^3} \mathcal H,
	\end{array}
\eeqs
where $P_0$ is a yet undetermined constant, the remaining BPS equations reduce to
\beqs
\label{eq:BPSeqs}
\begin{array}{rcl}
	\partial_y \mathcal A_J &=& \frac{v^2 \left(\mathcal B_J-\mathcal B_X\right)}{12 (v-2) (y-1)} , \ \ \
	\partial_y \mathcal B_J = \frac{\left(6 \mathcal A_J+ \mathcal B_J+ \mathcal B_X\right)}{(v-2) (y-1)} , \ \ \
	\partial_y \mathcal B_X = \frac{2 (v-2) \left(\mathcal B_J - 6 \mathcal A_J\right)}{v^2 (y-1) (y+1)^2} , \\[3mm]
	\partial_y \mathcal H &=& -\frac{P_0^3 v^3 (y+1) \left(12 \mathcal A_J \left(\mathcal B_J- \mathcal B_X\right)+\mathcal B_J (\mathcal B_J + 2\mathcal B_X) -3\right)}{36 P^3 (v-2)^2 (y-1)} .
	\end{array}
\eeqs
Contrary to the previous ones, these cannot be solved analytically, so we will resort to numerical calculations. In the following, we also fix
\beq
	\mathcal A_J = \frac{v^2+(v-1) v y-2}{6 v (y+2)-12} \mathcal B_J- \frac{v (v+1)(y+1)}{6 v (y+2)-12} \mathcal B_X ,
\eeq
which automatically solves Eq.~\eqref{eq:BPSeqs} for $\mathcal A_J$.

Depending on the range of the radial coordinate $y$, the backgrounds that we will study fall into two families, $\mathbb B_8^\pm$, parameterized by $y_0$. For the $\mathbb B_8^+$ family, the range of $y$ is given by $-1 < y_0 \leq y < 1$ with the IR located at $y = y_0$ and the UV at $y = 1$. These backgrounds have
\beqs
	v &=& \frac{1}{(1-y^2)^{1/4}} \left( v_0^+  + \,_2F_1\left[\frac12,\frac34;\frac32;y^2\right] \, y \right) , \ \ \ \
	P = P_0 \frac{(1+y)^{3/4}}{(1-y)^{1/4}} v ,
\eeqs
where $P_0 = \frac{Q_k^2(v_0^+ + v_c)}{4}$, $v_c \equiv \frac{\Gamma[1/4]^2}{\sqrt{8\pi}}$, and $v_0^+$ is fixed by the requirement that $v(y_0) = 2$.

For the $\mathbb B_8^-$ family, the range of $y$ is given by $1 < y \leq y_0 < \infty$ (as before the IR is located at $y = y_0$ and the UV at $y = 1$), and the backgrounds have
\beqs
	v &=& \frac{1}{(y^2-1)^{1/4}} \left( v_0^-  + \frac{2}{\sqrt{y}} \,_2F_1\left[\frac14,\frac34;\frac54;\frac{1}{y^2}\right] \right) , \ \ \
	P = P_0 \frac{(y+1)^{3/4}}{(y-1)^{1/4}} v ,
\eeqs
where $P_0 = \frac{Q_k^2(v_0^- + \sqrt{2} \, v_c)}{4}$ and $v_0^-$ is again fixed by the requirement that $v(y_0) = 2$.

In order to construct the backgrounds numerically, we set up the boundary conditions in the IR and evolve Eq.~\eqref{eq:BPSeqs} towards the UV, checking that we obtain the correct D2-brane asymptotics. It is convenient to work in the radial coordinate $\tau$ defined by
\beq
	y = \frac{1+y_0}{2} + \frac{1-y_0}{2} \tanh(\tau) ,
\eeq
in terms of which the IR (UV) is located at $\tau = -\infty$ ($\tau = +\infty$) for both $\mathbb B_8^\pm$ families. We define $\alpha \equiv e^{2\tau_*} \equiv 1 + y_0$, such that $-\infty < \tau_* < +\infty$. The IR expansions are given by
\beqs
	\mathcal B_J &=& 1 - \frac{1}{2\alpha} e^{2\tau} + \frac{\alpha +5}{8 \alpha ^2} e^{4\tau} -\frac{\alpha  (5 \alpha +22)+75}{80 \alpha ^3} e^{6\tau} + \mathcal O(e^{8\tau}) , \nonumber\\
	\mathcal B_X &=& 1 - \frac{3}{4\alpha^2} e^{4\tau} + \frac{\alpha +19}{8 \alpha ^3} e^{6\tau} -\frac{\alpha  (3 \alpha +34)+387}{64 \alpha ^4} e^{8\tau} + \mathcal O(e^{10\tau}) , \\
	\mathcal H &=& \mathcal H_{IR} -\frac{7 |2-\alpha|^{3/4}}{48 \alpha ^{13/4}} e^{2\tau} + \frac{7 (|2-\alpha|^{3/4} (\alpha +33)}{576 \alpha ^{17/4}} e^{4\tau} +\mathcal O(e^{6\tau}) .\nonumber
\eeqs
We determine the integration constant $\mathcal H_{IR}$ by requiring that $\mathcal H = 0$ in the UV, corresponding to the decoupling limit of the D2-branes. All in all, once the charges $Q_c$, $Q_k$ and $q_c$ are fixed, the backgrounds are labelled by a single parameter $\tau_*$. This means that once we select the gauge theory dual, with given gauge group and CS level, this parameter characterises completely the solution. As we already mentioned, it controls the difference between the microscopic couplings of each of the two factors in the gauge group.  

Finally, let us consider the solutions far in the UV, where the asymptotic expansions are given by ($z \equiv e^{-\tau/2}$)
\beqs
	\mathcal B_J &=& b_0 \left( 1+\frac{4 z}{\beta }+\frac{8 z^2}{\beta ^2}-\frac{16 z^3}{\beta ^3}+ \tilde{b}_4 z^4 \right) + \mathcal O(z^5) ,  \nonumber\\
	\mathcal B_X &=& b_0 \left( 1+\frac{4 z}{\beta }+\frac{12 z^2}{\beta ^2}+\frac{32 z^3}{\beta ^3}-\left(\frac{\tilde{b}_4}{2}+\frac{64}{\beta
   ^4}\right) z^4 \right) + \mathcal O(z^5) ,  \\ \nonumber
	\mathcal H &=& -\frac{2 \left(\beta  \left(b_0^2-1\right) P_0^3\right) z^5}{15 \gamma ^3}+\frac{8 \left(1-2 b_0^2\right) P_0^3 z^6}{9 \gamma
   ^3}+\frac{32 \left(9-28 b_0^2\right) P_0^3 z^7}{63 \beta  \gamma ^3}+ \mathcal O(z^8) .
\eeqs
For $\mathbb B_8^+$ backgrounds, 
\beq
\beta = \frac{v_0^+ + v_c}{(4 - 2\alpha)^{1/4}},\quad\quad\quad \gamma = \frac{2 P_0 (v_0^+ + v_c)}{(4 - 2\alpha)^{1/2}},
\eeq
while for $\mathbb B_8^-$ backgrounds, 
\beq
\beta = \frac{v_0^- + \sqrt{2} v_c}{(2\alpha - 4)^{1/4}}, \quad\quad\quad\gamma = \frac{2 P_0 (v_0^+ + \sqrt{2} v_c)}{(2\alpha - 4)^{1/2}}.
\eeq
The constants $b_0$ and $\tilde b_4$ are both determined in terms of $\tau_*$. Note that in the UV limit the four-dimensional metric becomes proportional to
\beq
\label{eq:HSVmetric}
	\dd s_4^2 \sim \frac{4(1-b_0^2)(4q_c^2 - 3Q_c Q_k)\beta}{15 z^{5/2}} \dd z^2 + \frac{\gamma ^2}{z^{7/2}} \dd x_{1,2}^2 \, ,
\eeq
which under a rescaling $x \rightarrow \lambda x$, $z \rightarrow \lambda^{2/3} z$ exhibits hyperscaling violation $\dd s_4^2 \rightarrow \lambda^{\theta} \dd s_4^2$ with hyperscaling violation coefficient $\theta = -1/3$.

\begin{figure}[t]
\begin{center}
\includegraphics[width=15cm]{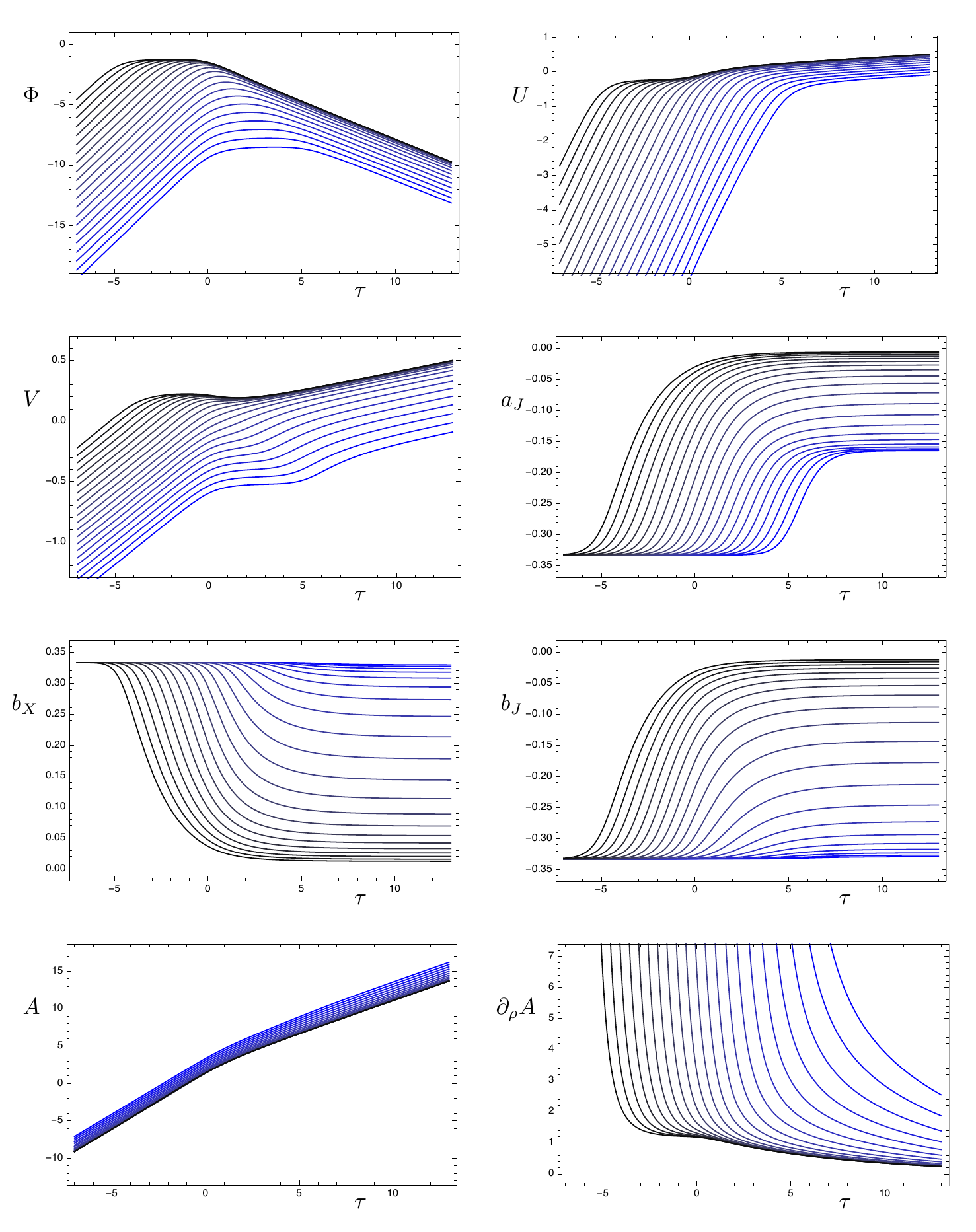}
\caption{Background functions for $\tau_* = -4.5, -4, \ldots, 5.5$. Lower (higher) values of $\tau_*$ correspond to black (blue). We have put $Q_c = 1/3$, $Q_k = -1$, $q_c = 0$ (the reason for this choice is explained in Section~\ref{sec:spectrum}).}
\label{Fig:backgrounds}
\end{center}
\end{figure}

Figure~\ref{Fig:backgrounds} shows the background functions for a few values of $\tau_*$. For large $\tau_*$ (corresponding to large $y_0$) the backgrounds approach that of the confining solution $\mathbb B_8^{\rm conf}$ described in Appendix~\ref{sec:B8conf}. For small $\tau_*$ (corresponding to $y_0$ approaching $-1$), the solutions flow close to the OP fixed point. Notice from the bottom-right panel that, for these solutions, $\partial_\rho A$ becomes nearly constant over a range of the radial coordinate, indicating that the backgrounds indeed are close to AdS in this region. A relevant deformation of the OP CFT makes these solutions exit the AdS region and develop a mass gap in the deep IR. This part of the RG flow is well approximated by the metric $\mathbb B_8^{\rm OP}$ \cite{Bryant:1989mv,Gibbons:1989er}. From the explicit form of the $\mathbb B_8^{\rm OP}$ solution given in Appendix~\ref{sec:B8OP}, it can be seen that the source for the relevant deformation is of the same order as any of the VEVs present.\footnote{The mode proportional to $\tilde{r}^{-1/3}$ in $\mathcal B_J$ and $\mathcal B_X$ corresponds to turning on a source for an operator of dimension $8/3$. The spectrum of scalars around the OP fixed point together with the dimension of the dual operators can be found in Table 1 of \cite{Faedo:2017fbv}.} This will be important for the physical interpretation of our results regarding the spectrum of composite states.

\subsection{Fluctuations}
\label{sec:formalism}

In this section, we summarize the gauge-invariant formalism that we will use to compute the spin-0 and spin-2 glueball spectra of the dual theory. We follow closely \cite{BHM1,LightScalars} to which the reader is referred for further details. 

Spectra are obtained by studying small fluctuations of the scalar fields and the metric around a given background. The fluctuations are taken to depend on both the radial coordinate $\rho$ as well the boundary coordinates $x^\mu$. After going to momentum-space (defining $m^2 = -q^\mu q^\nu \eta_{\mu\nu}$ where $q^\mu$ is the three-momentum and $\eta_{\mu\nu} = {\rm diag} (-1,1,1)$ is the boundary metric), one expands the equations of motion to linear order in the fluctuations and imposes the appropriate boundary conditions in the IR and UV. The spectrum is then given by those values of $m^2$ for which solutions exist.

More precisely, we expand in fluctuations $\{ \varphi^a, \nu, \nu^\mu, \mathfrak e^\mu{}_\nu, h, H, \epsilon^\mu \}$ as
\beqs
	\Phi^a &=& \bar \Phi^a + \varphi^a, \nonumber\\
	\dd s_4^2 &=& (1 + 2\nu + \nu_\sigma \nu^\sigma) \dd \rho^2 + 2 \nu_\mu \dd x^\mu \dd \rho + e^{2A} ( \eta_{\mu\nu} + h_{\mu\nu} ) \dd x^\mu \dd x^\nu, \\
	h^\mu_{\ \nu} &=& \mathfrak e^\mu_{\ \nu} + i q^\mu \epsilon_\nu + i q_\nu \epsilon^\mu + \frac{q^\mu q_\nu}{q^2} H + \frac{1}{2} \delta^\mu_{\ \nu} h,\nonumber
\eeqs
where $\mathfrak e^\mu_{\ \nu}$ is transverse and traceless, $\epsilon^\mu$ is transverse, and the three-dimensional indices $\mu$, $\nu$ are raised and lowered by the boundary metric $\eta$.

The spin-2 fluctuation $\mathfrak e^\mu_{\ \nu}$ satisfies the linearized equation of motion
\beq
\label{eq:fluceoms2}
	\left[\partial_\rho^2 + 3A' \partial_\rho + e^{-2A} m^2 \right] \mathfrak{e}^{\mu}_{\ \nu} = 0 .
\eeq
After forming the gauge-invariant combination \cite{BHM1,ESQCD}
\beq
	\mathfrak a^a = \varphi^a - \frac{\bar\Phi'^a}{4A'} h ,
\eeq
the linearized equation of motion for the spin-0 fluctuations can be written as
\beq
\label{eq:fluceoms}
	\Big[ \mathcal D_\rho^2 + 3A' \mathcal D_\rho + e^{-2A} m^2] \mathfrak{a}^a - \Big[ \mathcal V^a_{\ |c} - \mathcal{R}^a_{\ bcd} \bar \Phi'^b \bar \Phi'^d + \frac{2 (\bar \Phi'^a \mathcal V_c + \mathcal V^a \bar \Phi'_c )}{A'} + \frac{4 \mathcal V \bar \Phi'^a \bar \Phi'_c}{A'^2} \Big] \mathfrak{a}^c = 0.
\eeq
The different quantities involved in this expression are
\beqs
\begin{array}{rclcrcl}
\mathcal G_{abc} &=& \frac{1}{2} \left( \partial_b G_{ca} +\partial_c G_{ab} - \partial_a G_{bc} \right),&\qquad\qquad&\bar\Phi'_a &= &G_{ab} \partial_\rho\bar \Phi^b,\\[3mm]
\mathcal R^a_{\ bcd}& =& \partial_c \mathcal G^a_{\ bd} - \partial_d \mathcal G^a_{\ bc} + \mathcal G^a_{\ ce} \mathcal G^e_{\ bd} - \mathcal G^a_{\ de} \mathcal G^e_{\ bc},&\qquad\qquad&\mathcal V^a_{\ |b}& =& \frac{\partial \mathcal V^a}{\partial \Phi^b} + \mathcal G^a_{\ bc} \mathcal V^c,
\end{array}
\eeqs
that is, $\mathcal R^a_{\ bcd}$ is the Riemann tensor corresponding to the sigma model metric. On the other hand, the background covariant derivative is defined as $\mathcal D_\rho \mathfrak a^a = \partial_\rho \mathfrak a^a + \mathcal G^a_{\ bc} \bar \Phi'^b \mathfrak a^c$.

In order to obtain the spectrum, we first introduce IR (UV) cutoffs at $\rho_I$ ($\rho_U$). For backgrounds with an end-of-space in the IR located at $\rho = \rho_o$, the physical spectrum is obtained in the limit of $\rho_I \rightarrow \rho_o$, while similarly taking the limit of $\rho_U$ towards the location of the UV boundary. The boundary conditions at $\rho_I$ and $\rho_U$ are obtained by requiring that the variational problem be well-defined. This necessitates adding localized boundary actions in the IR and UV, which up to quadratic order in the fluctuations are determined by symmetry and consistency, except for a term quadratic in the scalar fluctuations. Taking the limit of this term corresponding to adding infinite boundary-localized mass terms for the scalar fluctuations, we obtain the boundary condition
\beq
	\varphi^a |_{\rho_{I,U}} = 0,
\eeq
which in terms of the gauge-invariant variable $\mathfrak a^a$ becomes
\beqs
\label{eq:BCs1}
	-\frac{e^{2A}}{m^2} \frac{\bar \Phi'^a}{A'} \left[ \bar \Phi'_b \mathcal D_\rho  - \frac{2\mathcal V \bar \Phi'_b}{A'} - \mathcal V_b \right] \mathfrak a^b \Big|_{\rho_i} = \mathfrak a^a \Big|_{\rho_{I,U}}.
\eeqs
Similarly, the boundary condition for the tensor fluctuations becomes
\beq
\label{eq:BCs2}
	\partial_\rho \mathfrak{e}^{\mu}_{\ \nu} |_{\rho_{I,U}} = 0 .
\eeq
These boundary conditions assure that subleading modes are selected in the limit of taking $\rho_I$ towards the end-of-space and $\rho_U$ towards the UV boundary, in accordance with standard gauge-gravity duality prescriptions.

Finally, we note that after a general change of radial coordinate from $\rho$ to $\tau$, Eq.~\eqref{eq:fluceoms} can be conveniently written as
\beqs
\label{eq:fluceomsnewr}
\Big[ \delta^a_{\ b}\partial_{\tau}^2+S^a_{\ b}\partial_{\tau}+T^a_{\ b}+ (\partial_{\tau} \rho)^2 e^{-2A}m^2 \delta^a_{\ b} \Big] \mathfrak{a}^b\,=\,0\,,
\label{Eq:ST}
\eeqs
where the matrices $S^a_{\ b}$ and $T^a_{\ b}$ are defined by
\beqs
\label{eq:SandT}
S^a_{\ b}&=&2{\cal G}^a_{\ bc}\partial_{\tau}\bar{\Phi}^c\,+\, \left[3\partial_{\tau}A - \partial_{\tau} \log(\partial_{\tau} \rho) \right] \delta^a_{\ b}\,,\\[2mm]
T^{a}_{\ b}&=&\partial_b{\cal G}^a_{\ cd}\partial_{\tau}\bar{\Phi}^c\partial_{\tau}\bar{\Phi}^d
- (\partial_{\tau} \rho)^2 \Bigg[ \left(\frac{2(\mathcal V^a\partial_{\tau}\bar{\Phi}^c+\mathcal V^c\partial_{\tau}\bar{\Phi}^a)}{\partial_{\tau}A}+
\frac{4\mathcal V\partial_{\tau}\bar{\Phi}^a\partial_{\tau}\bar{\Phi}^c}{(\partial_{\tau}A)^2}\right)G_{cb}\,+\partial_b \mathcal V^a \Bigg]\,.\nonumber
\eeqs

\section{Mass spectrum}
\label{sec:spectrum}

In this section, we numerically compute the spectrum of spin-0 and spin-2 glueballs in the theories dual to the backgrounds of Section~\ref{sec:backgrounds}. We start by making a number of points common to both the spin-0 and spin-2 calculations.

The sigma-model action $\mathcal S_4$ defined by Eq.~\eqref{eq:4daction}, Eq.~\eqref{eq:sigmametric}, and Eq.~\eqref{eq:superpotential} depends explicitly on the charges $Q_c$, $Q_k$, and $q_c$. However, after the redefinitions
\beqs
\begin{array}{rclcrclcrcl}
	\Phi &=& \tilde \Phi + \frac{1}{2} \log(X |Q_k|^{-3}) ,&\quad&U &=& \tilde U + \frac{1}{8} \log(X^3 |Q_k|^{-1}) , &\quad&V &=& \tilde V + \frac{1}{8} \log(X^3 |Q_k|^{-1}) , \\[3mm]
	A &=& \tilde A + \frac{1}{2} \log(X^3 |Q_k|^{-1}) ,&\quad&d \rho& =& \sqrt{X^3 |Q_k|^{-1}} d \tilde \rho , &\quad&	X& \equiv &\sqrt{4q_c^2 - 3 Q_c Q_k} ,
\end{array}
\eeqs
together with the analogous ones for $a_J$, $b_J$, and $b_X$ in Eq.~\eqref{eq:rescalings}, $\mathcal S_4$ factorizes into a functional of the fields and an overall factor that only depends on the charges. Since we are only interested in ratios of masses, we can therefore put in the numerical computations $Q_c = 1/3$, $Q_k = -1$ and $q_c = 0$ without loss of generality.

The second point concerns the boundary conditions used for the fluctuations. Following Section~\ref{sec:formalism}, we will impose Eq.~\eqref{eq:BCs1} and Eq.~\eqref{eq:BCs2} for the spin-0 and spin-2 modes, respectively, at an IR cutoff $\tau = \tau_I$, and make sure that the calculation of the spectrum converges to the physical result in the limit $\tau_I \rightarrow - \infty$. In principle, one should follow the analogous procedure in the UV, introducing a UV cutoff at $\tau = \tau_U$, and study the limit $\tau_U \rightarrow \infty$. However, we found that reaching high enough values of the UV cutoff is numerically challenging, and because of this we took a different approach, namely to make use of the explicit UV expansions of the fluctuations, selecting the subdominant modes (see Appendix~\ref{sec:UVexpansions}). As noted in Section~\ref{sec:formalism}, the two approaches agree in the limit $\tau_U \rightarrow +\infty$.

\subsection{Tensor modes}

In terms of the radial coordinate $y$, the equation of motion for the spin-2 fluctuations Eq.~\eqref{eq:fluceoms2}  becomes
\beq
	\Big[ \partial_y^2+\frac{v (y-2)-6 y}{2 (v-2) \left(y^2-1\right)} \partial_y + \frac{H P v}{4 (v-2) (y-1)^2 (y+1)} m^2 \Big] \mathfrak e^\mu_{\ \nu} = 0 .
\eeq
As promised, the dependence on $Q_k$, $Q_c$, $q_c$ enters only as a multiplicative factor in the term containing $m^2$ (through $H$ and $P$).

In order to understand the boundary conditions in the UV, we expand the general solution to Eq.~\eqref{eq:fluceoms2} in the UV as
\beqs
	\mathfrak e^\mu_{\ \nu}(\tau) &=& \tilde c^\mu_{\ \nu} \, \tilde{\mathfrak e}^{(UV)}(\tau) + c^\mu_{\ \nu} \, \mathfrak e^{(UV)}(\tau) ,\nonumber \\
	\tilde{\mathfrak e}^{(UV)}(\tau) &=& 1-\frac{2 \beta  \left(b_0^2-1\right) m^2 e^{-3 \tau /2}}{45 \gamma ^2}+\frac{4 \left(2-7 b_0^2\right) m^2 e^{-2 \tau }}{45 \gamma ^2}\nonumber\\
	&& -\frac{8 \left(21 b_0^2-1\right) m^2 e^{-5 \tau /2} \tau }{315
   \beta  \gamma ^2} +\mathcal O(e^{-3\tau}) , \\
	\mathfrak e^{(UV)}(\tau) &=& e^{-5 \tau /2} +\frac{20 e^{-3 \tau }}{3 \beta }+ \frac{240 e^{-7 \tau /2}}{7 \beta ^2}+\mathcal O(e^{-4\tau}),\nonumber
\eeqs
where $\tilde c^\mu_{\ \nu}$ and $c^\mu_{\ \nu}$ are constants associated with the dominant and subdominant modes, respectively. Imposing the boundary condition Eq.~\eqref{eq:BCs2} at finite cutoff $\tau = \tau_U$ leads to
\beq
	\tilde c^\mu_{\ \nu} = \frac{75 \gamma ^2 e^{-\tau_U}}{2 \beta  \left(b_0^2-1\right) m^2} c^\mu_{\ \nu} + \mathcal O(e^{-3\tau_U/2}),
\eeq
and hence in the limit of $\tau_U \rightarrow \infty$, the subdominant modes are automatically selected. In the numerical computation, we thus impose directly
\beq
\label{eq:BCflucs2UV}
	\mathfrak e^\mu_{\ \nu}(\tau_U) = c^\mu_{\ \nu} \, \mathfrak e^{(UV)}(\tau_U)
\eeq
as the boundary condition in the UV. Conversely, in the IR we impose Eq.~\eqref{eq:BCs2} at $\tau = \tau_I$. Next, we evolve Eq.~\eqref{eq:fluceoms2} from the IR and UV, and match the solutions (and their derivatives) at an intermediate value of $\tau = \left(\tau_I + \tau_U\right)/2$. The values of $m^2$ for which the solutions can be matched give us the spectrum. As a final step, we make sure that the spectrum converges as $\tau_I \rightarrow -\infty$ and $\tau_U \rightarrow +\infty$.

\begin{figure}[t]
\begin{center}
\includegraphics[width=10cm]{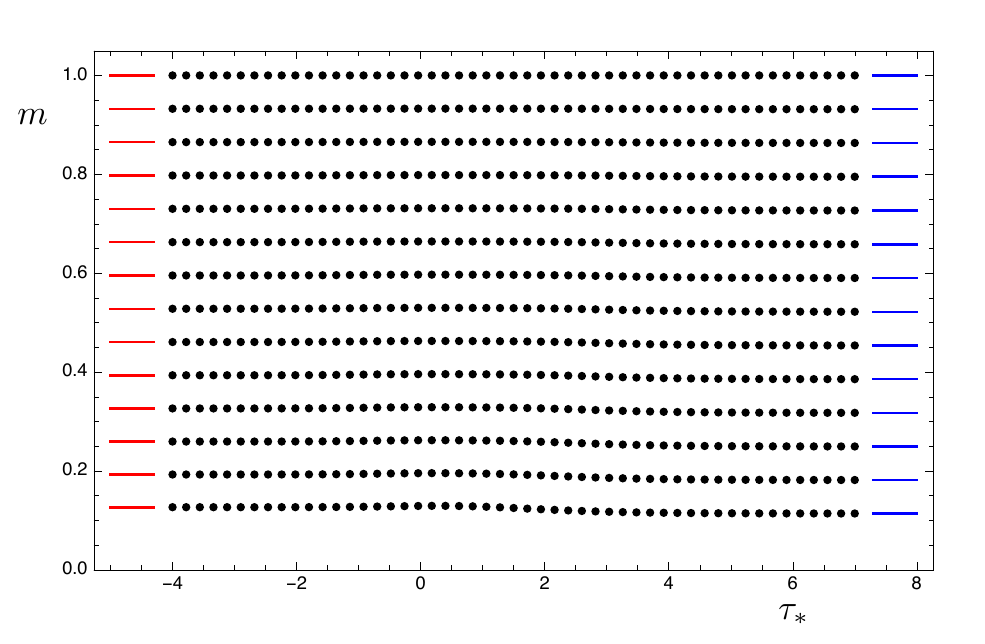}
\caption{Mass spectrum $m$ of spin-2 states as a function of $\tau_*$ normalized to the heaviest state included in the plot, compared to the spin-2 spectrum of $\mathbb B_8^{\rm OP}$ (red, left) and $\mathbb B_8^{\rm conf}$ (blue, right).}
\label{Fig:SpectrumSpin2}
\end{center}
\end{figure}

The resulting spectrum as a function of $\tau_*$ is shown in Figure~\ref{Fig:SpectrumSpin2}. For the numerics, we used $\tau_I = {\rm min}(-2,\tau_* - 6)$, $\tau_U = {\rm max}(5,\tau_* + 7)$. We have checked that these IR (UV) cutoffs are sufficiently low (high) for the spectrum to have converged to the physical result. As can be seen, there is a mass gap above which a tower of states appears. We have chosen to normalize the plot in such a way that the spacing of the heaviest modes remains the same as $\tau_*$ is varied in order to reflect the fact that the UV physics is the same for all the solutions. Conversely, the lowest lying states, which are sensitive to IR physics, show a slight dependence on $\tau_*$, becoming lighter as it is increased. In the limits of $\tau_* \rightarrow \pm\infty$, the spectra corresponding to the ${\mathbb B}_8^{\rm OP}$ and ${\mathbb B}_8^{\rm conf}$ backgrounds are recovered.

\begin{figure}[t]
\begin{center}
\includegraphics[width=15cm]{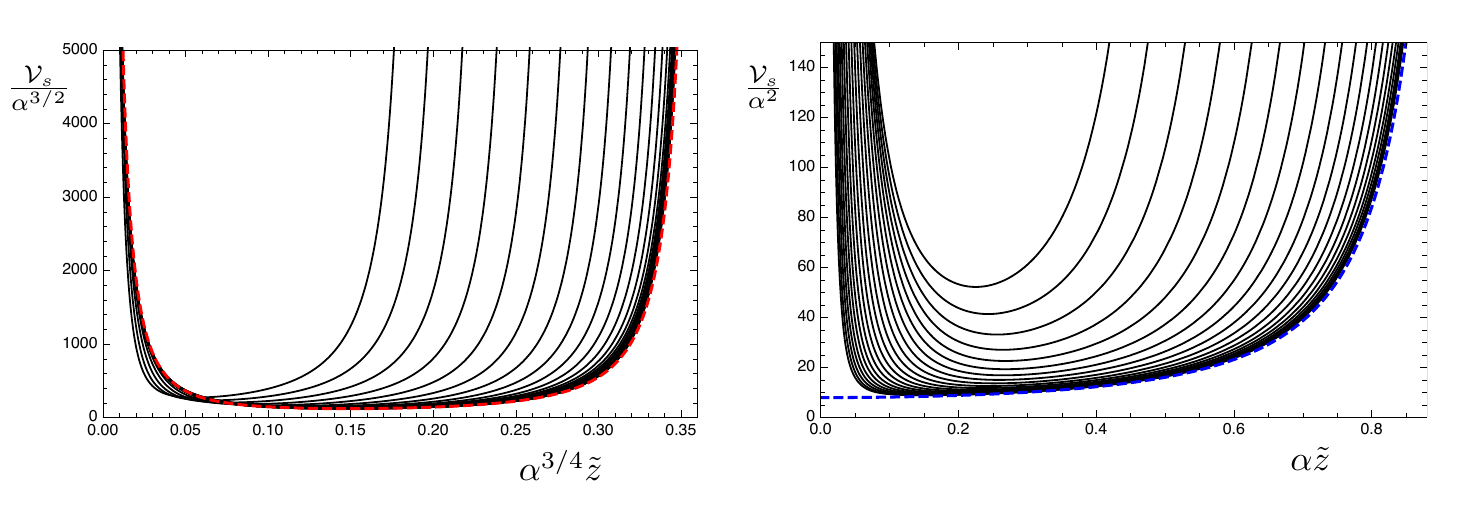}
\caption{The Schr\"odinger potential for a few different values of $\tau_*$ compared with the result for $\mathbb B_8^{\rm OP}$ (red) and $\mathbb B_8^{\rm conf}$ (blue).}
\label{Fig:Vs}
\end{center}
\end{figure}

In order to gain further intuition, let us rewrite Eq.~\eqref{eq:fluceoms2} so that it takes the Schr\"odinger form. After a change of radial coordinate $\dd\rho = e^A \dd\tilde z$ and a rescaling $\mathfrak e^\mu_{\ \nu} = e^{-A} e^{\mu}_{\ \nu}$, we obtain
\beq
	\Big( \partial_{\tilde z}^2 - \mathcal V_s(\tilde z) + m^2 \Big) e^{\mu}_{\ \nu} = 0, \qquad\quad\quad \mathcal V_s(\tilde z) = (\partial_{\tilde z} A)^2 + \partial_{\tilde z}^2 A .
\eeq
In Figure~\ref{Fig:Vs}, we plot the potential $\mathcal V_s$ as a function of ${\tilde z}$ for a few different backgrounds, compared to the result obtained for the $\mathbb B_8^{\rm conf}$ and $\mathbb B_8^{\rm OP}$. Note that to make this comparison in the $\tau_* \rightarrow \pm \infty$ limits, one has to take care to rescale $\mathcal V_s$ and $\tilde z$ by multiplying by $\alpha \equiv e^{2\tau_*}$ to the appropriate power (effectively choosing the units in which to measure).\footnote{See also Section~7 of \cite{Faedo:2017fbv}.} As can be seen, the resulting potentials are box-like, leading to a gapped and discrete spectrum.

\subsection{Scalar modes}

The scalar fluctuations satisfy the linearized equation of motion given in Eq.~\eqref{eq:fluceomsnewr}. The general solution can be expanded in the UV as
\beq
	\mathfrak a^a(\tau) = \tilde c_i \, \tilde{\mathfrak a}^{(UV)a}_i(\tau) + c_i \, \mathfrak a^{(UV)a}_i(\tau) \,.
\eeq
Here, $\tilde c_i$ and $c_i$ are constants ($i = 1, \cdots, 6$) associated with six dominant modes $\tilde{\mathfrak a}^{(UV)}_i(\tau)$ and six subdominant modes $\mathfrak a^{(UV)}_i(\tau)$, the explicit form of which is given in Appendix~\ref{sec:UVexpansions}. Out of the dominant modes, there is one that starts at order $z^{-5}$, one at order $z^{-4}$, two at order $z^0$, and two at order $z$, while the six subdominant modes start at orders $z^3$, $z^4$, $z^5$, $z^6$, $z^8$, and $z^{10}$, respectively. In order to select the subdominant modes, we hence impose the UV boundary condition
\beq
\label{eq:BCs1UV}
	\mathfrak a^a(\tau_U) = c_i \, \mathfrak a^{(UV)a}_i(\tau_U) \,.
\eeq

After incorporating the boundary conditions --- Eq.~\eqref{eq:BCs1} in the IR and Eq.~\eqref{eq:BCs1UV} in the UV --- we evolve Eq.~\eqref{eq:fluceomsnewr} from both sides and match at an intermediate value of $\tau$. In order to perform this matching, we use the midpoint determinant method \cite{BHM1}. More precisely, we form the
$12 \times 12$ matrix
\beq
	\mathcal M(\tau;m^2) =
	\left(
\begin{array}{cc}
 \mathfrak a^{(IR)}(\tau) & \mathfrak a^{(UV)}(\tau) \\
 \partial_\tau \mathfrak a^{(IR)}(\tau) & \partial_\tau \mathfrak a^{(UV)}(\tau)
\end{array}
\right) \, ,
\eeq
where $\mathfrak a^{(IR)}$ ($\mathfrak a^{(UV)}$) is a $6 \times 6$ matrix obtained by putting next to each other the column vectors corresponding to six linearly independent solutions that satisfy the boundary conditions in the IR (UV). The spectrum is given by those values of $m^2$ for which $\det \mathcal M(\tau;m^2) = 0$, in which case there exists a solution which can be written both as a linear combination of the solutions evolved from the IR and the UV. For the purpose of the numerics, we chose to evaluate $\det \mathcal M$ at an intermediate value of $\tau = (\tau_I + \tau_U)/2$. Finally, we make sure that the spectrum converges in the limits $\tau_I \rightarrow -\infty$ and $\tau_U \rightarrow \infty$.

\begin{figure}[t]
\begin{center}
\includegraphics[height=18cm]{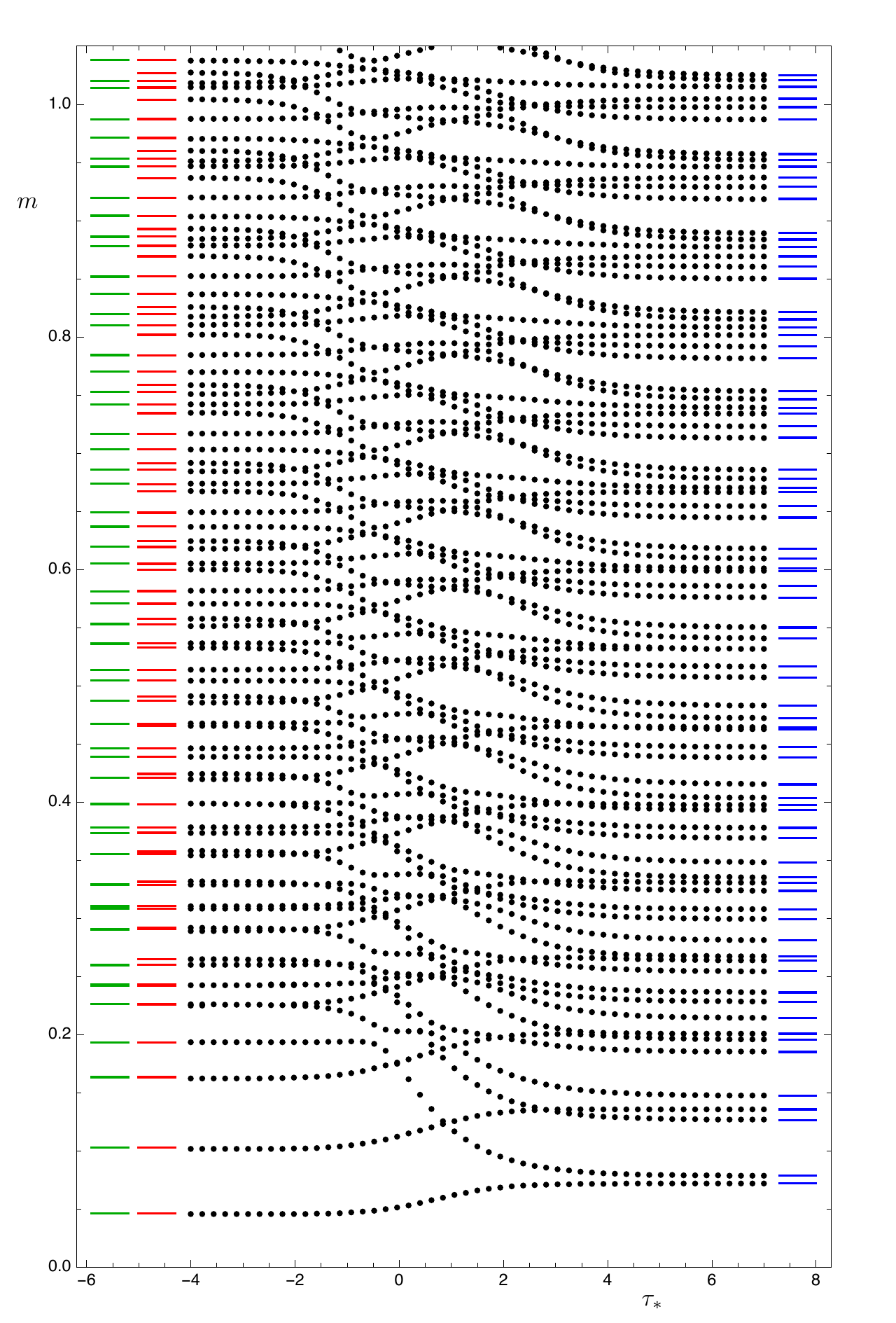}
\caption{Mass spectrum $m$ of spin-0 states as a function of $\tau_*$ with the same normalization as for Figure~\ref{Fig:SpectrumSpin2}. Also shown is the spin-0 spectrum of $\mathbb{B}_8^{\rm OP}$ (from left: in green for four scalars, in red for six scalars) and $\mathbb{B}_8^{\rm conf}$ (blue, right).}
\label{Fig:SpectrumSpin0}
\end{center}
\end{figure}

Figure~\ref{Fig:SpectrumSpin0} shows the scalar spectrum as a function of $\tau_*$. The normalization is the same as that for the spin-2 spectrum in Figure~\ref{Fig:SpectrumSpin2}, and we have also used the same values of the IR and UV cutoffs, namely $\tau_I = {\rm min}(-2,\tau_* - 6)$ and $\tau_U = {\rm max}(5,\tau_* + 7)$. For large values of $\tau_*$, the spectrum approaches that of the ${\mathbb B}_8^{\rm conf}$ background. It is interesting that the heavier states in this limit seem to come in groups of six. This may be due to the fact that these states are mostly sensitive to the UV D2-brane geometry, which is simpler than the full solution. Conversely, for small values of $\tau_*$, the spectrum approaches the one corresponding to the ${\mathbb B}_8^{\rm OP}$ solution. In this case, the theory flows close to the OP fixed point, and hence the ${\mathbb B}_8^{\rm OP}$ background is valid as an effective theory up to high energy scales. In this limit we observe that certain low-lying states become approximately degenerate. Presumably, this is due to an enhancement of symmetry that takes place in the IR of these flows. Indeed, the  $S^3$ of the internal geometry is not squashed in the ${\mathbb B}_8^{\rm OP}$ solution, which leads to an approximate $SO(4)$ symmetry enhancement in the IR of flows that pass very close to the OP fixed point. This same symmetry is the one that allows for a consistent truncation of the sigma model from six to four scalars that admits the ${\mathbb B}_8^{\rm OP}$ background as a solution. This truncation is described in Section 7.2 of \cite{Cassani:2011fu}, when the internal manifold is taken to be the seven-sphere.  Finally, we note that despite the fact that the theory can be made to flow arbitrarily close to the IR fixed point of OP as $\tau_* \rightarrow -\infty$, there is never a parametrically light state in the spectrum. We will explore this issue further in the following sections.

\subsection{Hard-wall cutoff in the IR}

In this section, we perform a numerical study of the scalar spectrum as a function of the IR cutoff $\tau_I$, focusing in particular on when a light state is present and which dynamics is responsible for lifting its mass. A similar study investigating cutoff effects was carried out in \cite{Athenodorou:2016ndx}, where it was argued that gauge-gravity duality can be used to facilitate the interpretation of lattice data.

\begin{figure}[t]
\begin{center}
\includegraphics[width=15cm]{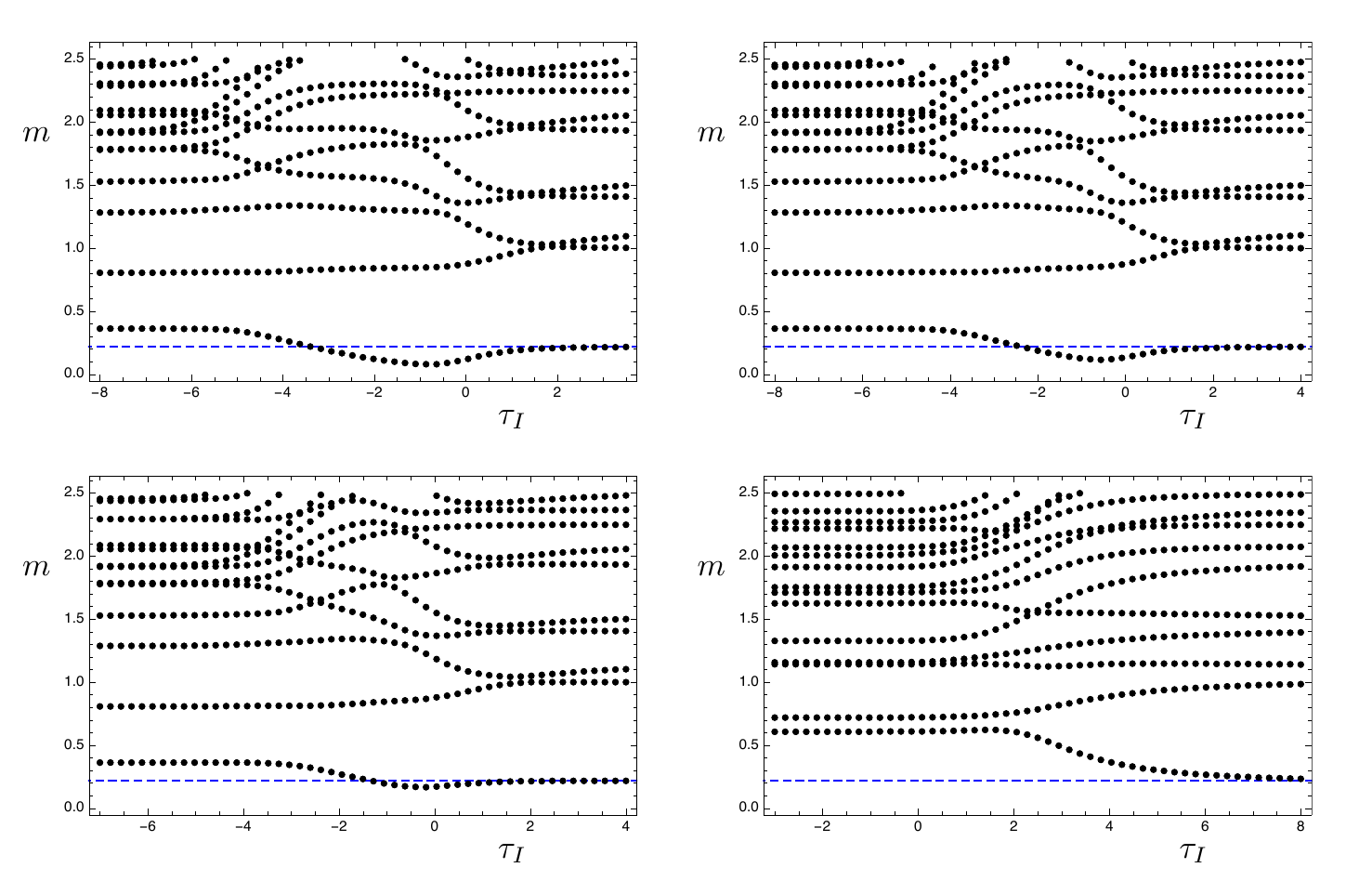}
\caption{Mass spectrum $m$ of spin-0 states as a function of the IR cutoff $\tau_I$, for a few backgrounds with different values of $\tau_*$. The top-left panel has $\tau_* = -4$, the top-right panel has $\tau_* = -3$, the bottom-left panel has $\tau_* = -2$, and the bottom-right panel has $\tau_* = 3$. The dashed blue line corresponds to the mass of the lightest state in the hyperscaling violating toy model of Appendix~\ref{sec:HSV}. All the plots are normalized to the mass of the lightest spin-2 state.}
\label{Fig:varyIR}
\end{center}
\end{figure}

Figure~\ref{Fig:varyIR} shows the dependence of the scalar spectrum as a function of $\tau_I$ for a few different backgrounds. Consider first the top-left panel, for which $\tau_* = -4$, hence describing a flow that comes close to the OP fixed point. As $\tau_I$ is varied, there are three different behaviours.
\begin{itemize}
	\item For small values of $\tau_I$, the spectrum coincides with the physical spectrum obtained in the limit $\tau_I \rightarrow -\infty$.
	\item As $\tau_I$ is increased such that the IR cutoff is within the region close to AdS, there is a state whose mass decreases and becomes light.
	\item Increasing $\tau_I$ further such that the IR cutoff is located within the far-UV region of the background, corresponding to the D2-branes, the mass of the light state again is lifted. However, it is still parametrically light compared to the mass of the second lightest state.
\end{itemize}

There are a number of examples in the literature analogous to the second case. In these examples an RG flow that passes near an IR fixed point is modelled by the introduction of a hard-wall in an AdS geometry, which typically leads to the presence of a light state in the spectrum \cite{PremKumar:2010as,LightScalars}. As can be seen in our model, when $\tau_*$ is increased, the effect becomes less pronounced. Indeed, in the bottom-right panel for which $\tau_* = 3$, describing a flow that is close to the ${\mathbb B}_8^{{\rm conf}}$ background, the intermediate region with a light state disappears altogether.

It may seem puzzling that in the third case, when $\tau_I$ is located in the far-UV region of the background, there is also a parametrically light state. This is due to the fact that in this region the four-dimensional metric exhibits hyperscaling violation with coefficient $\theta = -1/3$, as demonstrated by Eq.~\eqref{eq:HSVmetric}. Since AdS corresponds to $\theta = 0$, we expect that for small $|\theta|$ there will be a light dilaton which becomes exactly massless in the limit $\theta \rightarrow 0$. In Appendix~\ref{sec:HSV}, we study a simple toy model consisting of a single scalar field with a potential chosen in such a way that there exist hyperscaling violating solutions (see also \cite{Elander:2015asa}). Performing a perturbative analysis of the spin-0 spectrum, we find that there is a light state whose mass $m_0$ is given, to leading order in $\theta$, by
\beq
	\frac{m_0}{m_1} = \frac{1}{\pi} \sqrt{\frac{-3\theta}{2}} \,,
\eeq
where $m_1$ is the mass of the lightest spin-2 state. This estimate agrees well with the numerical result for the relevant case $\theta = -1/3$ (see Figure~\ref{Fig:spectrumHSV} of Appendix~\ref{sec:HSV}). As can be seen in Figure~\ref{Fig:varyIR}, for large $\tau_I$, the mass of the lightest spin-0 state approaches the value captured by this toy model (the dashed blue line).

\section{Conclusions and Discussion}
\label{sec:conclusions}

We performed a numerical study of the spectrum of spin-0 and spin-2 glueballs in a one-parameter family of three-dimensional theories by making use of their dual supergravity descriptions. Working within a consistent truncation to a sigma model coupled to four-dimensional gravity allowed us to utilize the gauge-invariant formalism to treat the scalar and tensor fluctuations in the bulk. As the parameter $\tau_*$ is varied, the spectrum interpolates between the two limiting cases corresponding to the confining solution $\mathbb B_8^{\rm conf}$ (for large $\tau_*$) and the solution given by $\mathbb B_8^{\rm OP}$ (for small $\tau_*$), the latter of which is dual to the OP CFT deformed by a relevant operator that induces a mass gap in the deep IR.

Despite the dual theory exhibiting quasi-conformal behaviour for backgrounds with small $\tau_*$, we did not find a light pseudo-dilaton in the physical spectrum. The RG flow described by such backgrounds consists of three parts. The first part takes the theory from the UV corresponding to the D2-branes down to the vicinity of the OP fixed point, following closely the solution $\mathbb B_8^{\infty}$. The second part consists of a region where the geometry is approximately AdS, and the dual theory remains close to the OP fixed point. From the point of view of this fixed point, how long it stays close is determined by the size of an irrelevant operator. The smaller $\tau_*$ is, the smaller this irrelevant operator is. The third and final part describes the flow from the OP fixed point to the deep IR where the theory develops a mass gap.

For small $\tau_*$, this final part of the RG flow is very well captured by the ${\mathbb B}_8^{\rm OP}$ solution, which describes the OP CFT with a relevant deformation that induces the flow towards an IR theory with a mass gap. The source of this relevant operator is of the same order as the VEVs present, and hence the ratio of explicit to spontaneous breaking of scale invariance is of order one, lifting the mass of any potential light dilaton. This interpretation is further supported by the study of the spectrum as a function of the location of a hard-wall IR cutoff. As long as this IR cutoff remains inside the AdS region corresponding to the OP fixed point, there is a light state (for small $\tau_*$) in the scalar spectrum. However, when the IR cutoff is taken to be close to the end-of-space, this state ceases to be light, hence showing that it is the physics of the deep IR that lifts its mass.

\vspace{1.0cm}
\begin{acknowledgments}

We were supported by grants FPA2016-76005-C2-1-P, FPA2016-76005-C2-2-P, SGR-2017-754, MDM-2014-0369 of ICCUB, and ERC Starting Grant HoloLHC-306605. DP was supported by fellowship FPA2013-46570-C2-2-P. JGS acknowledges support from the FPU program, fellowship FPU15/02551. DE is supported by the OCEVU Labex (ANR-11-LABX-0060) and the A*MIDEX project (ANR-11-IDEX-0001-02) funded by the ``Investissements d'Aveni'' French government program managed by the ANR.

\end{acknowledgments}

\appendix
\allowdisplaybreaks

\section{UV expansions of the scalar fluctuations}
\label{sec:UVexpansions}

We list here the UV expansion of the spin-0 fluctuations around the $\mathbb{B}_8^{\pm}$ backgrounds. The six dominant fluctuations are given by:

\beqs
\nonumber
\mathfrak {\tilde a}^{(UV)}_1 &=&
z^{-5}
\left(
\begin{array}{c}
 \frac{12}{5} \\
 1 \\
 1 \\
 0 \\
 0 \\
 0 \\
\end{array}
\right)
+ z^{-4}
\left(
\begin{array}{c}
 -\frac{16 \left(13 b_0^2-7\right)}{7 \beta  \left(b_0^2-1\right)} \\
 -\frac{4 \left(113 b_0^2-63\right)}{35 \beta  \left(b_0^2-1\right)} \\
 -\frac{8 \left(53 b_0^2-28\right)}{35 \beta  \left(b_0^2-1\right)} \\
 \frac{8 b_0 \left(7-13 b_0^2\right)}{21 \beta  \left(b_0^2-1\right)} \\
 -\frac{16 b_0}{15 \beta } \\
 \frac{16 b_0}{15 \beta } \\
\end{array}
\right)
+ z^{-3}
\left(
\begin{array}{c}
 \frac{64 \left(531 b_0^4-307 b_0^2+76\right)}{245 \beta ^2 \left(b_0^2-1\right){}^2} \\
 \frac{16 \left(2942 b_0^4-1969 b_0^2+527\right)}{735 \beta ^2 \left(b_0^2-1\right){}^2} \\
 \frac{8 \left(5317 b_0^4-3224 b_0^2+907\right)}{735 \beta ^2 \left(b_0^2-1\right){}^2} \\
 \frac{64 b_0 \left(37 b_0^2-7\right)}{105 \beta ^2 \left(b_0^2-1\right)} \\
 0 \\
 \frac{64 b_0 \left(7-22 b_0^2\right)}{105 \beta ^2 \left(b_0^2-1\right)} \\
\end{array}
\right)
+ \mathcal O(z^{-2}) , \\ \nonumber
\mathfrak {\tilde a}^{(UV)}_2 &=&
z^{-4}
\left(
\begin{array}{c}
 0 \\
 0 \\
 0 \\
 1 \\
 0 \\
 0 \\
\end{array}
\right)
+ z^{-3}
\left(
\begin{array}{c}
 0 \\
 0 \\
 0 \\
 -\frac{16}{3 \beta } \\
 \frac{8}{3 \beta } \\
 \frac{4}{3 \beta } \\
\end{array}
\right)
+ z^{-2}
\left(
\begin{array}{c}
 0 \\
 0 \\
 0 \\
 \frac{8}{\beta ^2} \\
 -\frac{8}{\beta ^2} \\
 -\frac{8}{\beta ^2} \\
\end{array}
\right)
+ z^{-1}
\left(
\begin{array}{c}
 0 \\
 0 \\
 0 \\
 -\frac{4 \beta  \left(b_0^2-1\right) m^2}{405 \gamma ^2} \\
 0 \\
 0 \\
\end{array}
\right)
+ \mathcal O(z^0) , \\ \nonumber
\mathfrak {\tilde a}^{(UV)}_3 &=&
\left(
\begin{array}{c}
 \frac{15-9 b_0^2}{4 b_0-4 b_0^3} \\
 \frac{21 b_0^2-3}{16 b_0-16 b_0^3} \\
 \frac{21 b_0^2-3}{16 b_0-16 b_0^3} \\
 \frac{1}{2} \\
 1 \\
 -1 \\
\end{array}
\right)
+ z
\left(
\begin{array}{c}
 \frac{60 \left(b_0^3+b_0\right)}{7 \beta  \left(b_0^2-1\right){}^2} \\
 0 \\
 \frac{75 \left(b_0^3+b_0\right)}{14 \beta  \left(b_0^2-1\right){}^2} \\
 -\frac{2 \left(53 b_0^4-31 b_0^2+28\right)}{35 \beta  \left(b_0^2-1\right){}^2} \\
 0 \\
 0 \\
\end{array}
\right)
+ z^2
\left(
\begin{array}{c}
 -\frac{3 \left(1093 b_0^6+7566 b_0^4+6459 b_0^2-1918\right)}{343 \beta ^2 b_0 \left(b_0^2-1\right){}^3} \\
 \frac{15703 b_0^6-334966 b_0^4+192713 b_0^2-17850}{1372 \beta ^2 b_0 \left(b_0^2-1\right){}^3} \\
 \frac{415 b_0^6+61346 b_0^4-95407 b_0^2+6846}{1372 \beta ^2 b_0 \left(b_0^2-1\right){}^3} \\
 -\frac{4 \left(681 b_0^6+3412 b_0^4-3317 b_0^2+224\right)}{105 \beta ^2 \left(b_0^2-1\right){}^3} \\
 0 \\
 \frac{2 \left(37 b_0^4-89 b_0^2+42\right)}{7 \beta ^2 \left(b_0^2-1\right){}^2} \\
\end{array}
\right)
+ \mathcal O(z^3) , \\ \nonumber
\mathfrak {\tilde a}^{(UV)}_4 &=&
\left(
\begin{array}{c}
 -20 \\
 1 \\
 1 \\
 0 \\
 0 \\
 0 \\
\end{array}
\right)
+ z
\left(
\begin{array}{c}
 \frac{320 b_0^2}{7 \beta -7 \beta  b_0^2} \\
 0 \\
 \frac{200 b_0^2}{7 \beta -7 \beta  b_0^2} \\
 \frac{32 b_0 \left(193 b_0^2-168\right)}{105 \beta  \left(b_0^2-1\right)} \\
 \frac{64 b_0}{3 \beta } \\
 -\frac{64 b_0}{3 \beta } \\
\end{array}
\right)
+ z^2
\left(
\begin{array}{c}
 \frac{16 \left(1093 b_0^4+3589 b_0^2+1918\right)}{343 \beta ^2 \left(b_0^2-1\right){}^2} \\
 -\frac{4 \left(125463 b_0^4-215513 b_0^2+17850\right)}{1029 \beta ^2 \left(b_0^2-1\right){}^2} \\
 \frac{4 \left(54465 b_0^4-47911 b_0^2+6846\right)}{1029 \beta ^2 \left(b_0^2-1\right){}^2} \\
 \frac{64 b_0 \left(51 b_0^4-97 b_0^2+546\right)}{315 \beta ^2 \left(b_0^2-1\right){}^2} \\
 0 \\
 \frac{32 b_0 \left(98-93 b_0^2\right)}{21 \beta ^2 \left(b_0^2-1\right)} \\
\end{array}
\right)
+ \mathcal O(z^3) , \\ \nonumber
\mathfrak {\tilde a}^{(UV)}_5 &=&
z
\left(
\begin{array}{c}
 0 \\
 1 \\
 -\frac{1}{2} \\
 -\frac{2 b_0}{15} \\
 0 \\
 0 \\
\end{array}
\right)
+ z^2
\left(
\begin{array}{c}
 \frac{5-125 b_0^2}{49 \beta -49 \beta  b_0^2} \\
 \frac{6297-17873 b_0^2}{588 \beta -588 \beta  b_0^2} \\
 \frac{3111-7999 b_0^2}{588 \beta  \left(b_0^2-1\right)} \\
 \frac{4 b_0 \left(303 b_0^2-163\right)}{45 \beta  \left(b_0^2-1\right)} \\
 0 \\
 \frac{2 b_0}{\beta } \\
\end{array}
\right) + \\ \nonumber
&& z^3
\left(
\begin{array}{c}
 \frac{2 \left(73309 b_0^4-64900 b_0^2+11151\right)}{3087 \beta ^2 \left(b_0^2-1\right){}^2} \\
 \frac{2457027 b_0^4-3116524 b_0^2+692097}{6174 \beta ^2 \left(b_0^2-1\right){}^2} \\
 \frac{-993105 b_0^4+1368572 b_0^2-342867}{6174 \beta ^2 \left(b_0^2-1\right){}^2} \\
 -\frac{4 b_0 \left(20580 \left(201 b_0^4-262 b_0^2+61\right) \log (z)+1651831 b_0^4-2522952 b_0^2+903721\right)}{46305 \beta ^2
   \left(b_0^2-1\right){}^2} \\
 \frac{8 b_0 \left(28531-50011 b_0^2\right)}{6615 \beta ^2 \left(b_0^2-1\right)} \\
 \frac{16 b_0 \left(9938 b_0^2-4343\right)}{6615 \beta ^2 \left(b_0^2-1\right)} \\
\end{array}
\right)
+ \mathcal O(z^4) , \\ \nonumber
\mathfrak {\tilde a}^{(UV)}_6 &=&
z
\left(
\begin{array}{c}
 0 \\
 0 \\
 0 \\
 1 \\
 0 \\
 0 \\
\end{array}
\right)
+ z^2
\left(
\begin{array}{c}
 0 \\
 \frac{20 b_0}{\beta -\beta  b_0^2} \\
 \frac{10 b_0}{\beta  \left(b_0^2-1\right)} \\
 \frac{36 b_0^2+4}{3 \beta -3 \beta  b_0^2} \\
 -\frac{2}{\beta } \\
 -\frac{1}{\beta } \\
\end{array}
\right)
+ z^3
\left(
\begin{array}{c}
 -\frac{20 b_0 \left(19 b_0^2+11\right)}{21 \beta ^2 \left(b_0^2-1\right){}^2} \\
 -\frac{25 b_0 \left(249 b_0^2-239\right)}{21 \beta ^2 \left(b_0^2-1\right){}^2} \\
 \frac{5 b_0 \left(519 b_0^2-569\right)}{21 \beta ^2 \left(b_0^2-1\right){}^2} \\
 \frac{8 \left(105 \left(21 b_0^4-22 b_0^2+1\right) \log (z)+412 b_0^4-294 b_0^2-68\right)}{63 \beta ^2 \left(b_0^2-1\right){}^2} \\
 \frac{4 \left(43 b_0^2+17\right)}{9 \beta ^2 \left(b_0^2-1\right)} \\
 -\frac{4 \left(31 b_0^2-1\right)}{9 \beta ^2 \left(b_0^2-1\right)} \\
\end{array}
\right)
+ \mathcal O(z^4) \, .
\eeqs

Next, we list the six subdominant fluctuations which we use in setting up the boundary conditions in the UV (in the numerics we retained up to and including order $z^{14}$):

\beqs
\nonumber
\mathfrak a^{(UV)}_1 &=&
z^3
\left(
\begin{array}{c}
 0 \\
 0 \\
 0 \\
 1 \\
 0 \\
 0 \\
\end{array}
\right)
+ z^4
\left(
\begin{array}{c}
 0 \\
 0 \\
 0 \\
 \frac{4}{\beta } \\
 -\frac{2}{\beta } \\
 -\frac{1}{\beta } \\
\end{array}
\right)
+ z^5
\left(
\begin{array}{c}
 -\frac{224 b_0}{5 \beta ^2 \left(b_0^2-1\right)} \\
 0 \\
 0 \\
 \frac{96}{5 \beta ^2} \\
 -\frac{68}{5 \beta ^2} \\
 -\frac{12}{5 \beta ^2} \\
\end{array}
\right)
+ \mathcal O(z^6) , \\ \nonumber
\mathfrak a^{(UV)}_2 &=&
z^4
\left(
\begin{array}{c}
 0 \\
 1 \\
 -\frac{1}{2} \\
 -b_0 \\
 0 \\
 0 \\
\end{array}
\right)
+ z^5
\left(
\begin{array}{c}
 \frac{72-104 b_0^2}{5 \beta -5 \beta  b_0^2} \\
 0 \\
 -\frac{1}{\beta } \\
 -\frac{64 b_0}{15 \beta } \\
 \frac{32 b_0}{15 \beta } \\
 \frac{28 b_0}{15 \beta } \\
\end{array}
\right)
+ z^6
\left(
\begin{array}{c}
 \frac{8 \left(72351 b_0^4-97291 b_0^2+18200\right)}{1617 \beta ^2 \left(b_0^2-1\right){}^2} \\
 \frac{2 \left(219619 b_0^4-245989 b_0^2-8442\right)}{8085 \beta ^2 \left(b_0^2-1\right){}^2} \\
 -\frac{2 \left(131732 b_0^4-166577 b_0^2+21609\right)}{8085 \beta ^2 \left(b_0^2-1\right){}^2} \\
 -\frac{32 b_0 \left(1367 b_0^4-2149 b_0^2+678\right)}{1575 \beta ^2 \left(b_0^2-1\right){}^2} \\
 0 \\
 \frac{64 b_0}{3 \beta ^2} \\
\end{array}
\right)
+ \mathcal O(z^7) , \\ \nonumber
\mathfrak a^{(UV)}_3 &=&
z^5
\left(
\begin{array}{c}
 -20 \\
 1 \\
 1 \\
 0 \\
 0 \\
 0 \\
\end{array}
\right)
+ z^6
\left(
\begin{array}{c}
 \frac{80 \left(385-159 b_0^2\right)}{231 \beta  \left(b_0^2-1\right)} \\
 \frac{4 \left(1447 b_0^2-385\right)}{231 \beta  \left(b_0^2-1\right)} \\
 -\frac{4 \left(401 b_0^2+385\right)}{231 \beta  \left(b_0^2-1\right)} \\
 \frac{64 b_0 \left(3-7 b_0^2\right)}{45 \beta  \left(b_0^2-1\right)} \\
 0 \\
 0 \\
\end{array}
\right)
+ z^7
\left(
\begin{array}{c}
 -\frac{256 \left(828 b_0^4-5681 b_0^2+4598\right)}{1617 \beta ^2 \left(b_0^2-1\right){}^2} \\
 \frac{64 \left(15614 b_0^4-17529 b_0^2+3190\right)}{4851 \beta ^2 \left(b_0^2-1\right){}^2} \\
 \frac{32 \left(-11171 b_0^4+8958 b_0^2+4763\right)}{4851 \beta ^2 \left(b_0^2-1\right){}^2} \\
 \frac{256 b_0 \left(787-1378 b_0^2\right)}{3465 \beta ^2 \left(b_0^2-1\right)} \\
 \frac{256 b_0 \left(103 b_0^2-330\right)}{3465 \beta ^2 \left(b_0^2-1\right)} \\
 \frac{256 b_0 \left(359 b_0^2+330\right)}{3465 \beta ^2 \left(b_0^2-1\right)} \\
\end{array}
\right)
+ \mathcal O(z^8) , \\ \nonumber
\mathfrak a^{(UV)}_4 &=&
z^6
\left(
\begin{array}{c}
 -\frac{255 b_0}{11 \left(b_0^2-1\right)} \\
 -\frac{249 b_0}{44 \left(b_0^2-1\right)} \\
 \frac{147 b_0}{44 \left(b_0^2-1\right)} \\
 \frac{8 b_0^2}{5 \left(b_0^2-1\right)} \\
 1 \\
 -1 \\
\end{array}
\right)
+ z^7
\left(
\begin{array}{c}
 \frac{120 b_0 \left(778-799 b_0^2\right)}{539 \beta  \left(b_0^2-1\right){}^2} \\
 \frac{6 b_0 \left(4318-4493 b_0^2\right)}{539 \beta  \left(b_0^2-1\right){}^2} \\
 \frac{6 b_0 \left(2269 b_0^2-2444\right)}{539 \beta  \left(b_0^2-1\right){}^2} \\
 \frac{8 \left(83 b_0^2+5\right)}{55 \beta  \left(b_0^2-1\right)} \\
 \frac{264-552 b_0^2}{55 \beta -55 \beta  b_0^2} \\
 -\frac{6 \left(169 b_0^2-33\right)}{55 \beta  \left(b_0^2-1\right)} \\
\end{array}
\right) + \\ \nonumber
&& z^8
\left(
\begin{array}{c}
 -\frac{24 b_0 \left(2042583 b_0^4-4008736 b_0^2+1911553\right)}{49049 \beta ^2 \left(b_0^2-1\right){}^3} \\
 -\frac{6 b_0 \left(2622797 b_0^4-5126704 b_0^2+2412907\right)}{49049 \beta ^2 \left(b_0^2-1\right){}^3} \\
 \frac{6 b_0 \left(1118759 b_0^4-2369148 b_0^2+1341389\right)}{49049 \beta ^2 \left(b_0^2-1\right){}^3} \\
 -\frac{4 \left(5981324 b_0^6-13926569 b_0^4+10018316 b_0^2-2346071\right)}{245245 \left(b_0^2-1\right){}^3 \beta
   ^2}-\frac{\left(b_0^2-1\right) \beta ^2 m^2}{270 \gamma ^2} \\
 \frac{8 \left(23054 b_0^4-29375 b_0^2+7021\right)}{2695 \beta ^2 \left(b_0^2-1\right){}^2} \\
 -\frac{32 \left(13775 b_0^4-14657 b_0^2+1057\right)}{2695 \beta ^2 \left(b_0^2-1\right){}^2} \\
\end{array}
\right)
+ \mathcal O(z^9) , \\ \nonumber
\mathfrak a^{(UV)}_5 &=&
z^8
\left(
\begin{array}{c}
 0 \\
 0 \\
 0 \\
 1 \\
 0 \\
 0 \\
\end{array}
\right)
+ z^9
\left(
\begin{array}{c}
 0 \\
 \frac{8 b_0}{\beta  \left(b_0^2-1\right)} \\
 \frac{4 b_0}{\beta -\beta  b_0^2} \\
 \frac{32-48 b_0^2}{3 \beta -3 \beta  b_0^2} \\
 \frac{8}{3 \beta } \\
 \frac{4}{3 \beta } \\
\end{array}
\right)
+ z^{10}
\left(
\begin{array}{c}
 -\frac{5888 b_0}{25 \beta ^2 \left(b_0^2-1\right)} \\
 0 \\
 -\frac{168 b_0}{\beta ^2 \left(b_0^2-1\right)} \\
 \frac{16 \left(1015 b_0^2-507\right)}{105 \beta ^2 \left(b_0^2-1\right)} \\
 \frac{32 \left(25 b_0^2-16\right)}{15 \beta ^2 \left(b_0^2-1\right)} \\
 \frac{16 \left(25 b_0^2-13\right)}{15 \beta ^2 \left(b_0^2-1\right)} \\
\end{array}
\right)
+ \mathcal O(z^{11}) , \\ \nonumber
\mathfrak a^{(UV)}_6 &=&
z^{10}
\left(
\begin{array}{c}
 \frac{12}{5} \\
 1 \\
 1 \\
 0 \\
 0 \\
 0 \\
\end{array}
\right)
+ z^{11}
\left(
\begin{array}{c}
 \frac{32 \left(146 b_0^2-77\right)}{77 \beta  \left(b_0^2-1\right)} \\
 \frac{8 \left(1186 b_0^2-627\right)}{385 \beta  \left(b_0^2-1\right)} \\
 \frac{8 \left(812 b_0^2-649\right)}{385 \beta  \left(b_0^2-1\right)} \\
 \frac{64 b_0 \left(151-184 b_0^2\right)}{5775 \beta  \left(b_0^2-1\right)} \\
 -\frac{512 b_0}{165 \beta } \\
 \frac{512 b_0}{165 \beta } \\
\end{array}
\right)
+ z^{12}
\left(
\begin{array}{c}
 \frac{128 \left(282706 b_0^4-378817 b_0^2+96591\right)}{45815 \beta ^2 \left(b_0^2-1\right){}^2} \\
 \frac{16 \left(2884061 b_0^4-3806734 b_0^2+927473\right)}{137445 \beta ^2 \left(b_0^2-1\right){}^2} \\
 \frac{8 \left(2785387 b_0^4-4761752 b_0^2+1985965\right)}{137445 \beta ^2 \left(b_0^2-1\right){}^2} \\
 -\frac{16 b_0 \left(58169 b_0^2-38611\right)}{17325 \beta ^2 \left(b_0^2-1\right)} \\
 \frac{64 b_0 \left(4387-6105 b_0^2\right)}{5775 \beta ^2 \left(b_0^2-1\right)} \\
 \frac{32 b_0 \left(13255 b_0^2-9093\right)}{5775 \beta ^2 \left(b_0^2-1\right)} \\
\end{array}
\right)
+ \mathcal O(z^{13}) \, .
\eeqs

Finally, we note that taking into account that the UV expansion of the sigma-model metric is given by
\beq
	G_{ab} =
 \left(
\begin{array}{cc}
 G_1 & 0 \\
 0 & G_2 \\
\end{array}
\right)
\eeq
where
\beq
\nonumber
	G_1 = 
\left(
\begin{array}{ccc}
 \frac{1}{4} & 0 & 0 \\
 0 & 2 & 2 \\
 0 & 2 & 6 \\
\end{array}
\right) \,, \ \
	G_2 = \frac{15\beta}{8(1-b_0^2)} {\rm diag}\left(\frac{8z}{\beta} + \mathcal O(z^2),z^{-1}+\mathcal O(z^0),2z^{-1}+\mathcal O(z^0)\right) \,,
\eeq
it can be shown that it is possible to arrange the dominant and subdominant modes into pairs $(\tilde{\mathfrak a}^{(UV)}_i,\mathfrak a^{(UV)}_j)$ such that $\tilde{\mathfrak a}^{(UV)a}_i G_{ab} \mathfrak a^{(UV)b}_j \sim z^5$, as expected in backgrounds with the hyperscaling violating UV asymptotics given by Eq.~\eqref{eq:HSVmetric}.

\section{$\mathbb B_8^{\rm OP}$ solution}
\label{sec:B8OP}

This solution is based on the metrics found in \cite{Bryant:1989mv,Gibbons:1989er}. After a change of radial coordinate
\beq
	\frac{\dd\rho}{\dd r} = 12 \sqrt[4]{\frac{3}{5}} H^{3/4} r^{5/4} |Q_k|^{7/4} \left[1-\left(\frac{r_0}{r}\right)^{5/3}\right]^{1/4} \, ,
\eeq
the $\mathbb B_8^{\rm OP}$ solution is given by ($\tilde r \equiv r/r_0$)
\beqs
\begin{array}{rclcrcl}
	e^{\frac{16U}{3}} &=& \frac{864}{125} 2^{1/3} H r_0^3\tilde r^3 |Q_k|^{7/3} \left(1-\tilde r^{-5/3}\right)^3, &\quad\quad&e^{4\Phi} &=& \frac{27 Hr_0^3 \tilde r^3}{125 |Q_k|^3} \left(1-\tilde r^{-5/3}\right)^3, \\[3mm]
	e^{\frac{16V}{3}} &=& \frac{108\ 2^{2/3} H r_0^3\tilde r^3 |Q_k|^{7/3}}{\sqrt[3]{5}} \left(1-\tilde r^{-5/3}\right)^{1/3}, &\quad\quad&
	e^{4A}& =& \frac{559872}{125} H r_0^7 \tilde r^7 |Q_k|^5 \left(1-\tilde r^{-5/3}\right)^3 \, .
\end{array}	
\eeqs
Factoring charges according to Eq.~\eqref{eq:rescalings}, the fluxes and $H$ read
\beqs
	\mathcal{B}_J\,&=&\,\frac{1}{\tilde r^{1/3}} \, \ \ \
	\mathcal{B}_X\,=\,\frac{6\tilde r^{5/3}-1}{5\tilde r^2}\,, \ \ \
\mathcal{H} = \frac{5}{243}\left[\frac{1}{\tilde r^2}-\frac{9}{\tilde r^{1/3}}-3\,\frac{\tilde r^{4/3}-\tilde r^{1/3}}{\tilde r^{5/3}-1}\right] + \\
 &&\frac{4\sqrt{2}}{81}\left(\sqrt{5+\sqrt{5}}\arctan\left[\frac{\sqrt{10+2\sqrt{5}}}{4\tilde r^{1/3}+1-\sqrt{5}}\right]+\sqrt{5-\sqrt{5}}\arctan\left[\frac{\sqrt{10-2\sqrt{5}}}{4\tilde r^{1/3}+1+\sqrt{5}}\right]\right) \,.\nonumber
\eeqs
We can put $r_0 = 1$ without loss of generality since it enters as an overall energy scale. Notice that there are no more scales in the solution, so all the VEVs and sources in the gauge theory dual are $\mathcal{O}(1)$ in these units. In particular $\mathcal{B}_J$ and $\mathcal{B}_X$ contain the source of a dimension $8/3$ operator. 

\section{$\mathbb B_8^{\rm conf}$ solution}
\label{sec:B8conf}
The $\mathbb B_8^{\rm conf}$ solution, found in \cite{Cvetic:2001ma}, has  $Q_k = 0$. After a change of radial coordinate
\beq
\frac{\dd \rho}{\dd \zeta} = \zeta^4_0\  \zeta^3\  H^{3/4}  \, ,
\eeq
it is given by
\beqs
\begin{array}{rclcrcl}
e^{4\Phi} &=& H \,, &\qquad\qquad&e^{2U} &= &\zeta_0^2\ H^{3/8} \left[ 1 -  \frac{1}{\zeta^4}  \right] \zeta^2 \,, \\[3mm]
e^{V} &=& \zeta_0\ \zeta\ H^{3/16} \,, &\qquad\qquad&e^{2A}& =&\zeta_0^6 \ H^{1/2} \left[ 1 -  \frac{1}{\zeta^4}   \right] \zeta^6 \,,
\end{array}
\eeqs
together with the fluxes and $H$ 
\beqs
b_J &=& \frac{Q_c}{4q_c}+\frac{2q_c}{3\zeta_0}\left[\frac{{\zeta}\sqrt{{\zeta}^4-1}-\left(3{\zeta}^4-1\right)U({\zeta})}{{\zeta}^4-1}\right]\,,\nonumber \\ 
b_X &=& -\frac{Q_c}{4q_c}-\frac{2q_c}{3\zeta_0}\left[\frac{{\zeta}\sqrt{{\zeta}^4-1}-\left(3{\zeta}^4-1\right)U({\zeta})}{{\zeta}^4}\right]\,, \\ \nonumber
H &=& \frac{128q_c^2}{9\zeta_0^6}\int_{\zeta}^\infty\left[\frac{2-3\sigma^4}{\sigma^3\left(\sigma^4-1\right)^2}+\frac{\left(4-9\sigma^4+9\sigma^8\right)U(\sigma)}{\sigma^4\left(\sigma^4-1\right)^{5/2}}+\frac{2\left(1-3\sigma^4\right)U(\sigma)^2}{\sigma^5\left(\sigma^4-1\right)^3}\right]\dd\sigma\,.
\eeqs
In these equations we have defined 
\beq
U({\zeta}) = \int_1^{\zeta}\left(\sigma^4-1\right)^{-1/2}\dd \sigma = i \left[F\left(\arcsin(\zeta) | -1 \right)- \frac{\Gamma\left[\frac{1}{4}\right]^2}{4\sqrt{2\pi}}\right] \,,
\eeq
with $F(\phi| m)$ the elliptic integral of the first kind.
We can put $\zeta_0 = 1$ without loss of generality since it enters as an overall energy scale.

\section{Hyperscaling violating toy model}
\label{sec:HSV}

Consider a ($d+1$)-dimensional model consisting of a single scalar field $\chi$ coupled to gravity with the action
\beqs
	S&=&\int d\rho \, d^d x \, \sqrt{-g} \left(\frac{R}{4}-\frac{1}{2} \partial_M \chi \partial^M \chi - \mathcal V(\chi) \right) ,
\eeqs
where
\beqs
	\dd s^2 &=& e^{2 \delta \chi(\rho)} \dd\rho^2 + e^{2A(\rho)} \dd x_{1,d-1}^2 \, ,\nonumber \\
	\mathcal V(\chi) &=& - \frac{(d-\theta)(d-\theta-1)}{4} e^{-2\delta  \chi} \, ,
\eeqs
and we have defined
\beq
	\theta = \frac{(d-1)^2\delta^2}{(d-1)\delta^2 - 2} \, .
\eeq
The model admits solutions
\beqs
	\chi &=& \frac{\delta}{2} (d - \theta - 1) \rho \, , \nonumber\\
	A &=& \frac{d - \theta - 1}{d - 1} \rho \, .
\eeqs
Under a scale transformation $x \rightarrow \lambda x$, $\rho \rightarrow \rho - \log \lambda$, the metric transforms as
\beq
	\dd s^2 \rightarrow \lambda^{\frac{2\theta}{d-1}} \dd s^2 \, .
\eeq
This shows that the solutions are hyperscaling violating with hyperscaling violation coefficient $\theta$.

For $\theta$ close to $0$, the background metric is close to AdS, and because of this there should be a light state in the spectrum. In order to understand how its mass depends on $\theta$, we now perform a perturbative analysis. The equation of motion and boundary condition for the spin-0 fluctuation are given by
\beqs
\label{eq:eombcHSV}
	\partial^2_\rho \mathfrak a + (d - \theta) \partial_\rho \mathfrak a + e^{-2\rho} m^2 \mathfrak a &=& 0 \, ,\nonumber \\
	\frac{e^{2\rho}}{d - 1} \frac{\theta}{m^2} \partial_\rho \mathfrak a |_{\rho_{I,U}}&=& \mathfrak a |_{\rho_{I,U}} \, .
\eeqs
Consider the expansion (perturbative in $\theta$)
\beqs
	\mathfrak a &=& \mathfrak a_0 + \theta \, \mathfrak a_2 + \mathcal O(\theta^2) \, , \nonumber\\
	m^2 &=& \theta \, \tilde m^2 + \mathcal O(\theta^2) \, .
\eeqs
At leading order in $\theta$, the general solution of the equation of motion in Eq.~\eqref{eq:eombcHSV} is given by
\beq
	\mathfrak a_0 = c_1 + c_2 e^{-d \, \rho} \, ,
\eeq
where $c_1$ and $c_2$ are constants. After imposing the boundary conditions in the IR (UV), and solving for $\tilde m^2$, we obtain that the mass $m_0$ of the lightest spin-0 state is given by
\beq
	m_0^2 = \frac{d \, \theta}{d-1} \frac{e^{d \, \rho_U} - e^{2\rho_U}}{e^{d \, \rho_U} - 1} \, ,
\eeq
where we have fixed $\rho_I = 0$ without loss of generality. Finally, as long as $d > 2$, the limit of $\rho_U \rightarrow \infty$ leads to
\beq
\label{eq:md}
	m_0^2 = \frac{d \, \theta}{1 - d} \, .
\eeq

\begin{figure}[t]
\begin{center}
\includegraphics[width=10cm]{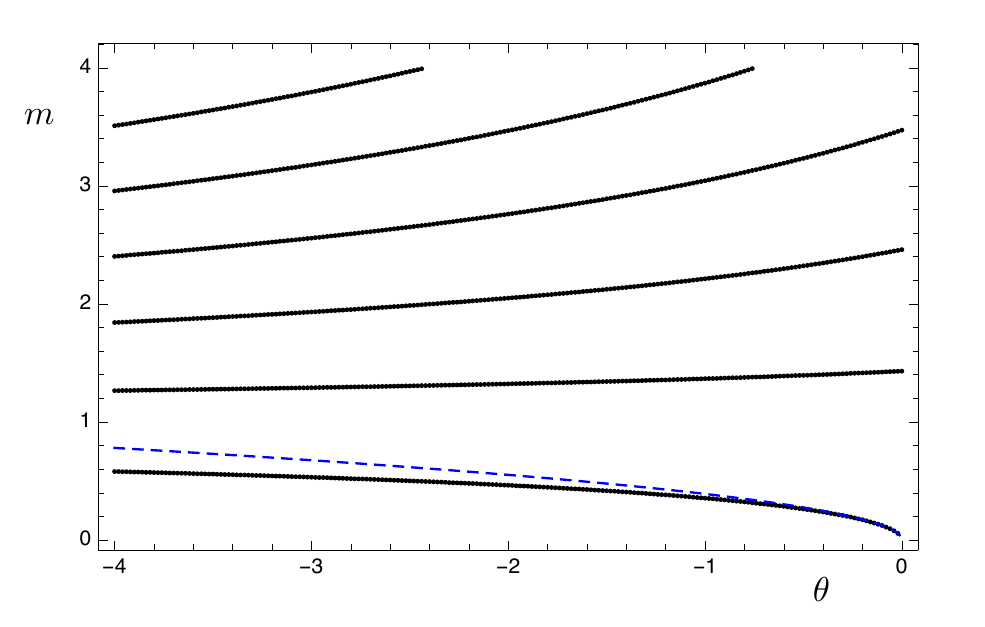}
\caption{Mass spectrum of spin-0 states normalized to the mass of the lightest spin-2 state for the hyperscaling violating toy model with $d = 3$, $\rho_I = 0$, $\rho_U = 50$. The dashed blue line is the perturbative result to leading order in $\theta$ given in Eq.~\eqref{eq:perturbativemassratio}.}
\label{Fig:spectrumHSV}
\end{center}
\end{figure}

The spin-2 fluctuations satisfy the same equation of motion as the spin-0 fluctuations
\beq
	\partial^2_\rho \mathfrak e^\mu_{\ \nu} + (d - \theta) \partial_\rho \mathfrak e^\mu_{\ \nu} + e^{-2\rho} m^2 \mathfrak e^\mu_{\ \nu} = 0 \, ,
\eeq
the general solution of which is given by
\beq
	\mathfrak e^\mu_{\ \nu} = e^{\frac{(\theta - d) \rho}{2}} m^{d/2} \left( c^\mu_{\ \nu} J_{\frac{d - \theta}{2}} (e^{-\rho} m) + d^\mu_{\ \nu} J_{\frac{\theta - d}{2}} (e^{-\rho} m) \right) \, ,
\eeq
where $c^\mu_{\ \nu}$ and $d^\mu_{\ \nu}$ are constants. After imposing the boundary condition
\beq
	\partial_\rho \mathfrak e^\mu_{\ \nu} |_{\rho_{I,U}} = 0 \, ,
\eeq
and taking the limit $\rho_U \rightarrow \infty$, one can show that for $\theta < d - 2$, the spin-2 states have masses given by the positive roots of the equation
\beq
	J_{\frac{d - \theta - 2}{2}} (m) = 0 \, .
\eeq
For $d=3$, one obtains that to leading order in $\theta$, the ratio between the masses of the lightest spin-0 and spin-2 states is given by
\beq
\label{eq:perturbativemassratio}
\frac{m_0}{m_1} = \frac{1}{\pi} \sqrt{\frac{-3\theta}{2}}
\eeq
In Figure~\ref{Fig:spectrumHSV}, we show the spin-0 spectrum for $d = 3$ as a function of $\theta$ compared to the perturbative result.

\end{document}